\documentclass[fleqn,usenatbib]{mnras}

\usepackage{newtxtext,newtxmath}

\usepackage[T1]{fontenc}

\DeclareRobustCommand{\VAN}[3]{#2}
\let\VANthebibliography\thebibliography
\def\thebibliography{\DeclareRobustCommand{\VAN}[3]{##3}\VANthebibliography}

\usepackage{graphicx}	

\newcommand{\Lsun}{\,$L_{\odot}$\,}	
\newcommand{\Msun}{\,$M_{\odot}$\,}	
\newcommand{\Rsun}{\,$R_{\odot}$\,}	
\newcommand{\Prot}{\,$P_{\rm rot}$\,}	
\newcommand{\vsini}{\,$v \sin(i)$\,}	
\newcommand{\Teff}{\,$T_{\rm eff}$\,}	
\newcommand{\mps}{m\,s\,$^{-1}$} 
\newcommand{\logg}{\,$\log(g)$\,} 
\newcommand{\Halpha}{\,H$\alpha$\,} %
\newcommand{\Mjup}{\mbox{M$_{\rm Jup}$}}
\newcommand{\gtsimeq}{\raisebox{-0.6ex}{$\,\stackrel{\raisebox{-.2ex}{$\textstyle >$}}{\sim}\,$}}

\title[TWA 7 and TWA 25 ]{The Surface Magnetic Activity of the Weak-Line T Tauri Stars TWA 7 and TWA 25}

\author[B. A. Nicholson et al.]{
B. A. Nicholson,$^{1,2}$\thanks{E-mail: belinda.nicholson@physics.ox.ac.uk}
G. Hussain,$^{3,4,5}$
J. -F. Donati,$^{5}$
D. Wright,$^{2}$
C. P. Folsom,$^{5,6}$\and
R. Wittenmyer,$^{2}$
J. Okumura,$^{2}$
B. D. Carter,$^{2}$
and the MaTYSSE collaboration
\\
$^{1}$Sub-department of Astrophysics, University of Oxford, Keble Rd, Oxford, United Kingdom\\
$^{2}$University of Southern Queensland, Centre for Astrophysics, Toowoomba, Australia\\
$^{3}$European Southern Observatory, Karl Schwarzschild Str. 2, 85748 Garching, Germany\\
$^{4}$Science Division, Directorate of Science, European Space Research and Technology Centre (ESA/ESTEC), Keplerlaan 1, 2201AZ, Noordwijk, The Netherlands\\
$^{5}$IRAP, Universit{\'{e}} de Toulouse, CNRS, UPS, CNES, 31400, Toulouse, France \\
$^{6}$Dept. of Physics \& Space Science, Royal Military College of Canada, PO Box 17000 Station Forces, Kingston, ON, Canada K7K 0C6 
}

\date{Accepted XXX. Received YYY; in original form ZZZ}

\pubyear{2021}

\begin{document}
\label{firstpage}
\pagerange{\pageref{firstpage}--\pageref{lastpage}}
\maketitle

\begin{abstract}
We present an analysis of spectropolarimetric observations of the low-mass weak-line T Tauri stars TWA 25 and TWA 7. The large-scale surface magnetic fields have been reconstructed for both stars using the technique of Zeeman Doppler imaging. Our surface maps reveal predominantly toroidal and non-axisymmetric fields for both stars. These maps reinforce the wide range of surface magnetic fields that have been recovered, particularly in  pre-main sequence stars that have stopped accreting from the (now depleted) central regions of their discs. We reconstruct the large scale surface brightness distributions for both stars, and use these reconstructions to filter out the activity-induced radial velocity jitter, reducing the RMS of the radial velocity variations from 495 \mps\, to 32 \mps\ for TWA 25, and from 127 \mps\ to 36 \mps\ for TWA 7, ruling out the presence of close-in giant planets for both stars. The TWA 7 radial velocities provide an example of a case where the activity-induced radial velocity variations mimic a Keplerian signal that is uncorrelated with the spectral activity indices. This shows the usefulness of longitudinal magnetic field measurements in identifying activity-induced radial velocity variations.
\end{abstract}


\section{Introduction}
Magnetic fields play a key role in the evolution of low mass stars onto the pre-main sequence (PMS), particularly during the T Tauri stage of cool star evolution. The T Tauri stage is divided into two major categories: classical T Tauri stars (cTTSs), and weak-line T Tauri stars (wTTSs). The cTTS stage starts when the central star has emerged from the cocoon of gas in which it formed, and is surrounded by a substantial disc of gas and dust, from which matter is accreted onto the stellar surface. At this time, magnetic fields are thought to help dissipate the angular momentum of the star and accreting matter such that the stellar rotation does not reach break-up speed as the star forms. Once the gas from the inner disc is cleared and accretion onto the star has ceased, the star is then classed as a wTTS; a fully formed star that is still undergoing contraction towards its main sequence size, often still surrounded by an extended debris disc. 

The study of the differences between these populations of PMS stars has been one of the goals of the Magnetic Topologies of Young Stars and Survival of close in Giant Exoplanets (MaTYSSE) large program. MaTYSSE is a multi-telescope programme using high resolution spectropolarimetric observations to map the large scale brightness and magnetic fields of a range of wTTSs, and compare them to a sample of cTTSs observed in the Magnetic Protostars and Planets\footnote{\url{https://wiki.lam.fr/mapp/FrontPage}} (MaPP) program \citep[see e.g.][]{Donati2007, Donati2008, Donati2010, Donati2011, Donati2012, Hussain2009}. 

Few stars below $\sim0.8$ \Msun have been observed in either sample, as their peak emission in the infrared and low brightness make them challenging targets for current optical spectropolarimeters. Within the MaPP sample there have been 3 PMS stars below this limit (when compared with the \cite{Baraffe2015} PMS evolution models), DN Tau \citep[0.65 \Msun;][]{Donati2013}, BP Tau \citep[0.7 \Msun;][]{Donati2008}, and V2247 Oph \citep[0.35 \Msun;][]{Donati2010}, and two stars with masses around 0.8 \Msun, TW Hya \citep{Donati2011b} and AA Tau \citep{Donati2010b}. The magnetic field of V2247 Oph appeared significantly different {compared to the other stars’ magnetic field}. {The star} is fully convective like its higher mass counterparts of the same age, but displayed a wildly different magnetic field morphology.

To further our understanding of low mass PMS dynamo fields, this work presents an analysis of high-resolution spectropolarimetric data of two low-mass wTTS stars as part of the MaTYSSE sample: TWA 25 and TWA 7. Both are 10 Myr old stars in the TW Hya association \citep{Mentuch2008}, and represent more evolved versions of the MaPP stars AA Tau (in the case of TWA 25), and DN Tau (in the case of TWA 7). 

In this analysis we reconstruct the surface brightness and large-scale magnetic field morphologies of both stars. In addition, we examine the variation in longitudinal (line of sight) magnetic field with stellar rotation, as well as the spectral activity markers of \Halpha and Na I doublet indices. 

The other major focus of the MaTYSSE program is the search for close-in giant planets around wTTSs. Finding planets around very active stars is challenging, as the surface activity of these stars can contribute to radial velocity variations of the order of hundreds of meters per second, obscuring even a giant planet's signal. The MaTYSSE program has addressed this by using the surface activity information provided from surface brightness mapping to measure and remove the radial velocity contribution due to this activity \citep{Donati2014}. This work, therefore, also analyses the radial velocities of TWA 25 and TWA 7, and uses the surface brightness information from our Doppler mapping to filter out activity jitter from the radial velocities of both stars. 

In Section \ref{sec:obs}, we detail the observations and data processing of TWA 25 and 7. The evolutionary states of both stars are determined, and the stellar properties calculated in Section \ref{sec:starpar}. Section \ref{sec:TomMod} describes the mapping of large-scale brightness and magnetic fields, Section \ref{sec:Blongs} explores the rotational modulation of the longitudinal magnetic field and activity indices, and Section \ref{sec:radvel} describes our analysis of the radial velocities for both stars. Lastly, we summarise and discuss our findings in Section \ref{sec:summary}.

\section{Observations}
\label{sec:obs}
High-resolution spectropolarimetric data were taken for TWA 25 and TWA 7 using  HARPS in polarimetric mode on the ESO 3.6-m telescope in La Silla, Chile. TWA 25 was observed {16 times} {and TWA 7 observed 17 times } from March 12$^{\rm th}$ to 31$^{\rm st}$, 2017. Journals of the observations for each star are given in Tables \ref{tab:obs_twa25} and \ref{tab:obs_twa7}.  Each observation consists of 4 sub-exposures with alternating configurations of a quarter wave plate to remove first order spurious polarisation signals. Two observations of TWA 7 contain only two sub-exposures due to weather, and so are only used for the analysis involving brightness information, and are excluded from the magnetic field analysis. Our spectra have a wavelength coverage of $\sim$380 nm to $\sim$690nm, and spectral resolution of $\sim$115000. The spectral data were reduced using the {\sc Libre ESpRIT} pipeline software adapted for use with HARPS polarimetric data \citep{Hebrard2016}, following the procedure outlined in \cite{Donati1997}. Across our observations we attain a peak circular polarisation (Stokes V) signal-to-noise ratio (SNR) of between 72 and {119} for TWA 25, with a median value of 95,  and between 55 and 110 for TWA 7, with a median value of 92.

\begin{table*}
	\centering
	\caption{Journal of observations for star TWA 25. From left to right, this table lists the date of observation, the Heliocentric Julian date,  the exposure time as a set of sub-exposures, the stellar rotational phase based on a period of 5.07 days (with the zero-point set as the middle of the first observation), the peak SNR in the Stokes {I and} V spectra per observation, and the SNR in Stokes {I and} V after least squares deconvolution (LSD, see Section \ref{sec:LSD}). {The final column gives the status of magnetic field detection based on False Alarm Probably (FAP) values, with definite detection (D) for FAP<$10^{-5}$,  marginal detection (M) for $10^{-5}<$FAP$<10^{-3}$, and non-detection (N) for FAP$>10^{-3}$.  }}
	\label{tab:obs_twa25}
	\begin{tabular}{ccccccccccc} 
		\hline
		Date   & HJD         & Exposure Time (s) & Rotation & { Stokes I}  & Stokes V  & {Stokes I} & Stokes V & Detection\\
		(2017) & (2457000+)  &                   & Cycle    & { Spec. SNR} & Spec. SNR & {LSD SNR}  & LSD SNR  & Status\\
		\hline
		 12 March & 825.83150 & $4\times1600$ & 0.00 & {107} & 103  & {637 } & {4381} & {D} \\
		 13 March & 826.82463 & $4\times1600$ & 0.20 & {78} & 74   & {617 } & {2907} & {D} \\
		 16 March & 829.80554 & $4\times1600$ & 0.78 & {120} & 116  & {653 } & {4809} & {D} \\
		 17 March & 830.83104 & $4\times1600$ & 0.99 & {99} & 92   & {635}  & {3812} & {D} \\
		 18 March & 831.81629 & $4\times1600$ & 1.18 & {108}   & 102 & {618} &{4088} & {D} \\
 		 20 March & 833.81797 & $4\times1600$ & 1.58 & {84} & 83  &  { 654 } & {3304} & {D} \\
		 21 March & 834.86501 & $4\times1600$ & 1.78 & {113} & 111  & {653 } & {4424} & {D} \\	
		 22 March & 835.86077 & $4\times1600$ & 1.98 & {99} & 95  & {633 } & {3727} & {D} \\
		 23 March & 836.82937 & $4\times1600$ & 2.17 & {93} & 91  & {616 } & {3494} & {D} \\
		 24 March & 837.82846 & $4\times1600$ & 2.37 & {91} & 88  & {627 } & {3433} & {D} \\
		 25 March & 838.82821 & $4\times1600$ & 2.56 & {93} & 91  & {642 } & {3583} & {D} \\
		 26 March & 839.82925 & $4\times1600$ & 2.76 & {123} & 119 & {651 } & {4838} & {D} \\
		 27 March & 840.83804 & $4\times1600$ & 2.96 & {100} & 95  & {639} & {3832} & {D} \\
		 28 March & 841.82974 & $4\times1600$ & 3.16 & {75} & 72  & {607} & {2779} & {D} \\
		 30 March & 843.82681 & $4\times1600$ & 3.55 & {108} & 104 & {640} & {4203} & {D} \\
		 31 March & 844.83717 & $4\times1600$ & 3.75 & {109} & 105 & {656} & {4241} & {D} \\
		\hline
	\end{tabular}
\end{table*}

\begin{table*}
	\centering
	\caption{Table of observations for star TWA 7. For description, see caption of Table \ref{tab:obs_twa25}. Rotation cycle is calculated based on a 5.00 day period. {The observations on the 18th and 19th March were incomplete due to weather, and so are excluded from the magnetic field analysis.}}
	\label{tab:obs_twa7}
	\begin{tabular}{ccccccccccc} 
		\hline
		Date   & HJD         & Exposure Time (s) & Rotation &Stokes I  & Stokes V  & Stokes I & Stokes V & Detection\\
		(2017) & (2457000+)  &                   & Cycle    &Spec. SNR & Spec. SNR & LSD SNR  & LSD SNR  & Status\\
		\hline
		 12 March & 825.75835 & $4\times1300$ & 0.00 & 74 & 73  & 396  & 2450 & N \\
		 13 March & 826.74890 & $4\times1500$ & 0.20 & 75 & 71  & 392  & 2535 & N \\
		 16 March & 829.69035 & $4\times1600$ & 0.79 & 111 & 110 & 382  & 3945 & D \\
		 17 March & 830.75377 & $4\times1600$ & 1.00 & 96 & 92  & 393  & 3279 & N \\
		 18 March & 831.72533 & $2\times1600$ & 1.19 & 75 & 72  & 404  & 2665 & - \\
 		 19 March & 832.72506 & $2\times1600$ & 1.39 & 55 & 55  & 360 & 2060 & - \\
		 21 March & 834.78337 & $4\times1650$ & 1.80 & 86 & 82  & 390 & 2877 & D \\		 
		 22 March & 835.78202 & $4\times1650$ & 2.00 & 75 & 71  &  391  & 2468 & N \\
		 23 March & 836.74918 & $4\times1750$ & 2.20 & 90 & 87  & 388  & 3044 & M \\
		 24 March & 837.74774 & $4\times1750$ & 2.40 & 107 & 101 & 389  & 3727 & D \\
		 25 March & 838.53731 & $4\times1750$ & 2.56 & 111 & 106 & 391  & 3993 & D\\
		 25 March & 838.74812 & $4\times1750$ & 2.60 & 97 & 92  & 382 & 3296 & D \\
		 26 March & 839.74884 & $4\times1750$ & 2.80 & 106 & 101  & 376 & 3600 & D \\
		 27 March & 840.75775 & $4\times1750$ & 3.00 & 101 & 98  & 372 & 3410 & N \\
		 28 March & 841.74953 & $4\times1750$ & 3.20 & 78 & 74  & 381 & 2612 & M \\
		 30 March & 843.74606 & $4\times1750$ & 3.60 & 103 & 101 & 386 & 3593 & M \\
		 31 March & 844.75641 & $4\times1750$ & 3.80 & 105 & 105 & 386 & 3622 & D \\
		\hline
	\end{tabular}
\end{table*}

\subsection{Least Squares Deconvolution}
\label{sec:LSD}
\begin{figure}
  \centering
    \includegraphics[width=.4\textwidth]{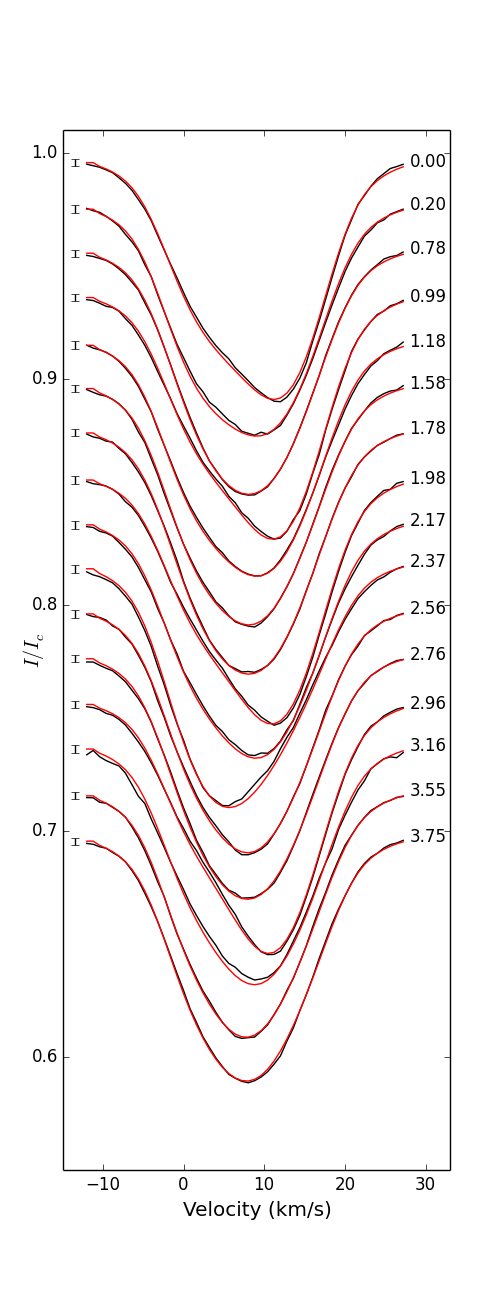}
  \caption{Stokes I LSD profiles (black line) of TWA 25, and fits to these profiles using {\sc DOTS} (red line, see Section \ref{sec:DI}). Mean $1\sigma$ error bars are given to the left of each profile, and rotation phase is given on the right.  }
  \label{fig:TWA25_final_fit}
\end{figure}

\begin{figure}
  \centering
    \includegraphics[width=.4\textwidth]{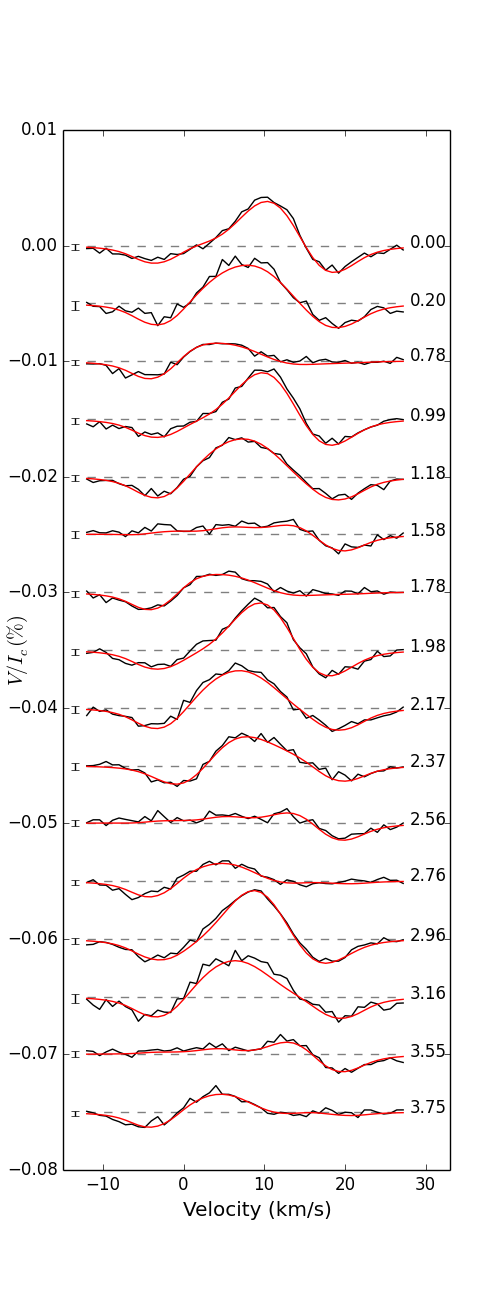}
  \caption{Circular polarisation (Stokes V) LSD profiles for TWA 25 (black line), with fits from Zeeman Doppler Imaging (red line, see Section \ref{sec:ZDI}). Mean error bars for each observation are shown on the left hand side, and rotation phase of each observation is given on the right hand side. }
  \label{fig:TWA_StokesV_fit_fin}
\end{figure}

\begin{figure}
  \centering
    \includegraphics[width=.4\textwidth]{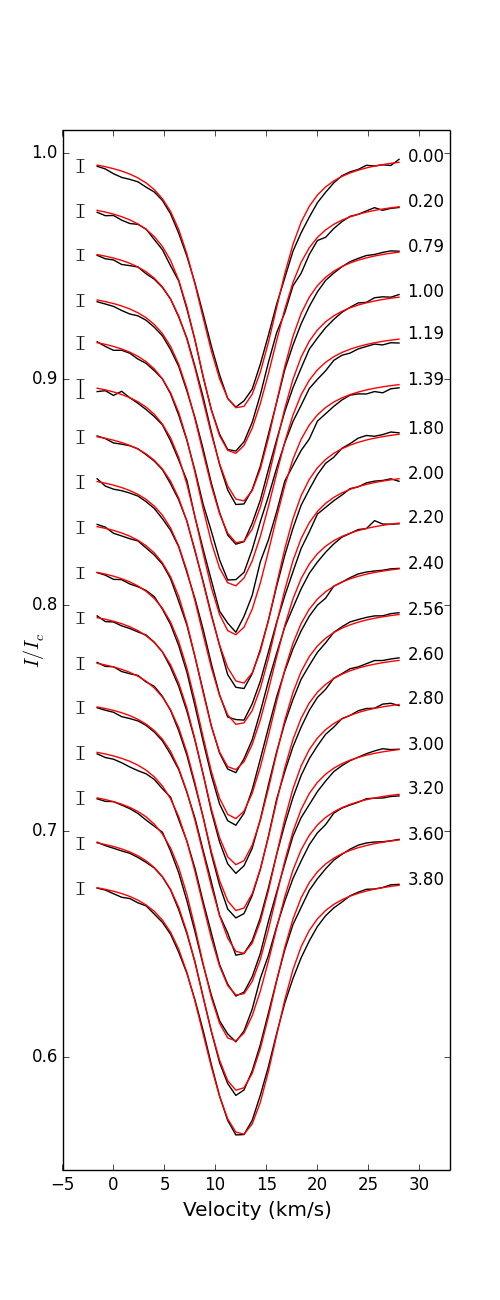}
  \caption{Stokes I LSD profiles (black line) of TWA 7, and fits to these profiles using {\sc DOTS} (red line, see Section \ref{sec:DI}). Mean $1\sigma$ error bars are given on the left-hand side of each profile, and rotation phase is given on the right. }
  \label{fig:TWA7_final_fit}
\end{figure}

\begin{figure}
  \centering
    \includegraphics[width=.4\textwidth]{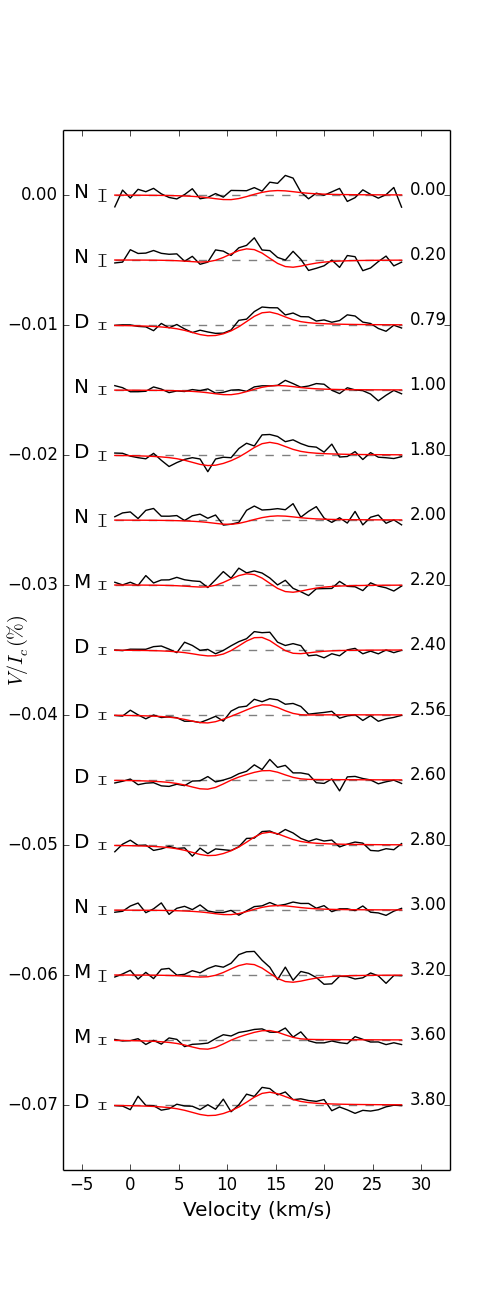}
  \caption{Circular polarisation (Stokes V) LSD profiles for TWA 7 (black line), with fits from Zeeman Doppler Imaging (red line, see Section \ref{sec:ZDI}). On the left hand side of each profile is the mean error bar, and a letter denoting the level of magnetic signal detection for each observation are shown on the left hand side, and rotation phase of each observation is given on the right hand side. The observations taken on March 18 and 19 were incomplete and are not shown in this plot as they are excluded from the magnetic field analysis.}
  \label{fig:TWA7_StokesV_fit_rvtweak2}
\end{figure}

To increase the signal-to-noise for our tomographic mapping, we combine the lines in our observed spectra using Least Squares Deconvolution \citep[LSD,][]{Donati1997}. This process combines the signal in absorption features, by deconvolving our observed spectra with a mask of photospheric lines constructed using the Vienna Atomic Line Database \citep[VALD3,][]{Ryabchikova2015} based on the effective temperature and surface gravity of each star (see Section \ref{sec:starpar}), and removing lines in emission, or with broad absorption. Given the large number of weak and blended absorption lines at the spectral types of these stars, we remove lines with a strength relative to the deepest lines of less than 0.1, as per \cite{Nicholson2018} to limit the effect of blends. After generating these LSD profiles, the continuum levels were re-normalised, and all were scaled by the mean equivalent width for each star using the re-normalisation program within the {\sc ZDIPy} package \citep{Folsom2018}. The Stokes I and Stokes V LSD profiles are plotted as black lines in Figures \ref{fig:TWA25_final_fit} and \ref{fig:TWA_StokesV_fit_fin} for TWA 25, and \ref{fig:TWA7_final_fit} and \ref{fig:TWA7_StokesV_fit_rvtweak2} for TWA 7. For TWA 25, the peak SNR in the intensity spectra increase from 123 to 656 in the Stokes I LSD profiles, and in the circular polarisation spectra the peak SNR increased from 119 to 4838 in the Stokes V LSD profiles. For TWA 7, we obtain an increase in peak SNR of our intensity spectra from 111 to 404 in the LSD profiles, and in the circular polarisation spectra we get an increase from 110 to 3993 in the Stokes V LSD profiles. 

\section{Stellar Parameters of TWA 25 and TWA 7}
\label{sec:starpar}
TWA 25 and TWA 7 are both single stars in the TW Hya association \citep{Song2003,Webb1999}, which has a mean age of 10 Myrs \citep{Mentuch2008}. To estimate the individual ages, and hence the evolutionary states for each star, we estimate their effective temperatures and surface gravities from observations, and calculate bolometric luminosities based on properties of the stars found in the literature. A summary of the stellar parameters discussed in this section and used for our analysis is given in Table \ref{tab:stellar_param}. 

\begin{table}
	\centering
	\caption{Summary of the calculated properties of TWA 25 and 7.}
	\label{tab:stellar_param}
	\begin{tabular}{lcc}
		\hline
		 & TWA 25 & TWA 7\\
		\hline
		Mass, $M_*$ (\Msun) & $0.83^{+0.04}_{-0.06}$ & $0.62\pm0.03$ \\   
		Radius, $R_*$ ($R_{\odot}$) & $1.2\pm0.2$ & $0.8^{+0.2}_{-0.1}$ \\  
		Age (Myrs) & $7^{+7}_{-4}$ & $17^{+19}_{-9}$ \\
		Luminosity, $L_*$ (\Lsun) & $0.4\pm0.1$ & $0.13^{+0.05}_{-0.03}$ \\
		Distance, (pc) & $53.1\pm0.2$ & $34.03\pm0.08$ \\
		\vsini (k\mps) & $11.9\pm0.3$ & $4.5\pm0.2$ \\
		Rotation Period, \Prot (Days) & $5.07\pm0.03$ & $5.00\pm0.01$ \\
		Effective Temperature, \Teff (K) & $4120\pm50$ & $3800\pm50$\\
		Surface Gravity, $\log(g)$ & $4.25\pm0.1$ & $4.7\pm0.2$\\
		Inclination Angle, $i$ (Degrees) & $57\pm5$  & $40^{+5}_{-15}$ \\
		\hline
	\end{tabular}
\end{table}

\subsection{Effective temperature and surface gravity}
We calculate the effective temperature (\Teff) and logarithmic surface gravity (\logg) for both stars using the HARPS observations presented here, following the method outlined in \cite{Donati2012}, which is based on the procedure of \cite{Valenti2005}. This method determines effective temperatures and surface gravities by comparing select atomic absorption regions in high resolution optical spectra to a grid of template synthetic spectra.  For TWA 25 we find an \Teff $ = 4120 \pm 50$ K and  \logg $= 4.3\pm0.1$, and for TWA 7 we find \Teff $ =3800 \pm 50$ K and  \logg $= 4.7\pm0.2$. The larger uncertainty on \logg for TWA 7 is due to the solution being at the edge of the model grid. 

There are many \Teff values for TWA 25 and TWA 7 within the literature, ranging from 3742 K \citep{daSilva2009} to 4250 K \citep{Ammons2008} for TWA 25, and from 3300 K \citep{Yang2008} to 4017 K \citep{GaiaDR22018} for TWA 7. Such a wide range in effective temperatures is unsurprising for pre-main sequence stars, and is likely due to the differences in methods used and the relative sensitivity of those methods to the presence of photospheric spots. For example, both \cite{Ammons2008} and \cite{GaiaDR22018} are large surveys, performing the same temperature analysis across their whole sample. Such a broad-brush approach is typically not appropriate for PMS stars due to their atypical nature. 

\cite{Mentuch2008} perform a detailed spectral analysis consistently for both stars, determining both \Teff and \logg values, as is needed for pre-main sequence stars. Mentuch et al. analyse high-resolution spectra of both stars using select atomic and molecular absorption regions, and find effective temperatures and logarithmic surface gravities of $3920\pm150$ and $4.45\pm0.5$ for TWA 25, and $ 3540\pm150$ and $4.18\pm0.5$ for TWA 7. Whilst the surface gravity estimates are in agreement within the uncertainties, the effective temperatures differ by more than 1 sigma of either estimate.  This discrepancy in effective temperature estimate is likely due to the inclusion of molecular absorption regions, which will give a systematically lower temperature due to those lines also being present in cool spots on the stellar surface. This is also the likely explanation for the lower temperature estimates of \cite{daSilva2009} for TWA25 and \cite{Yang2008} for TWA 7.

\subsection{Luminosity}
Luminosity is calculated for both stars based on their apparent magnitudes, extinction estimates, bolometric corrections and distances. For TWA 25 we use an apparent V magnitude of $11.160\pm0.083$ mag from \cite{Henden2016}, and a bolometric correction of $-0.97\pm0.05$ mag from \cite{Pecaut2013}. Interstellar extinction, $A_v$, is calculated using the observed B-V colour of $1.428\pm0.162$ mag from \cite{Henden2016} and intrinsic colour, $(B-V)_0$, of $1.18\pm0.03$ mag from \cite{Pecaut2013}, giving $A_v=0.8\pm0.5$ mag for TWA 25. The distance to TWA 25 is calculated as $53.1\pm0.2$ pc from parallax measurements from \textit{Gaia} Data Release 2 \citep[DR2,][]{GaiaCollab2016,GaiaDR22018}. These values result in a luminosity of $0.4^{+0.2}_{-0.1}$ \Lsun for TWA 25.  This value is higher than any previously published values, though the large uncertainty means that it is consistent with literature values. Our luminosity value is in closest agreement with the value of $0.252$ \Lsun published by \cite{McDonald2017}, who estimate an even higher extinction value of 0.643 mag, but use a smaller distance derived from the first \textit{Gaia} data release. The extinction values for TWA 25 are all calculated (here and in the literature) as greater than zero despite its close proximity (making significant interstellar extinction unlikely), indicating reddening due to spots on the stellar surface. 

To calculate the luminosity of TWA 7 we use an apparent V magnitude of $11.754\pm0.57$ mag from \cite{Henden2016}, and a bolometric correction of $-1.34\pm0.08$ mag from \cite{Pecaut2013}. We calculate $A_v$ using observed B-V colour of $1.475\pm0.063$ mag from \cite{Henden2016} and (B-V)$_0$ of $1.37\pm0.02$ from \cite{Pecaut2013}, giving $A_v=0.2\pm0.2$ mag for TWA 7.  Our estimate of extinction, given its large uncertainties, is equivalent to the usual assumed extinction of zero for this star. The \textit{Gaia} DR2 parallax measurement of TWA 7 gives a distance of $34.03\pm0.08$ pc. These values result in a luminosity of $0.08\pm 0.02$ \Lsun for TWA 7. This value is also in good agreement with all literature values, with the exception of a very high estimate of 0.32 \Lsun by \cite{Low2005}, who calculate a stellar luminosity by integrating a spectral energy distribution constructed from \textit{Spitzer} data, assuming an averaged distance to TWA of 55 pc. Scaling this luminosity value to the updated \cite{GaiaDR22018} distance of 34.03 pc we get a luminosity estimate of $\sim 0.12$ \Lsun, which is in better agreement with the other values in the literature. 

Both stars are expected to have heavily spotted photospheres that influence their observed colours and visual magnitudes, and in turn their luminosity estimates. Given stellar evolution models do not account for spotted photospheres, we apply a first-order spot correction to our luminosity values to make better estimates of the stars' evolutionary states. Following the method of \cite{Yu2019}, we assume $A_v=0$ (a reasonable assumption for stars this close) and an average spot coverage of $50 \pm 15 \%$, chosen as a conservative estimate based on the finding of total spot coverage of V140 Tau \citep[$50-75\%$][]{Yu2019} and LkCa 4\citep[$\sim 80\%$][]{Gully2017}. Although our tomographic mapping results (see Section \ref{sec:DI}) give a smaller percentage spot coverage, DI is not sensitive to the small scale spot features that contribute to the reddening, and thus underestimate the total area of spot coverage. With this assumed extinction and spot coverage values, we obtain a spot-adjusted luminosity of $0.4\pm0.1$ \Lsun for TWA 25 and $0.12^{+0.04}_{-0.03}$ \Lsun for TWA 7. 

\subsection{Masses, Radii, Ages and Evolutionary stages}
\begin{figure}
 \centering
 \includegraphics[width=0.48\textwidth]{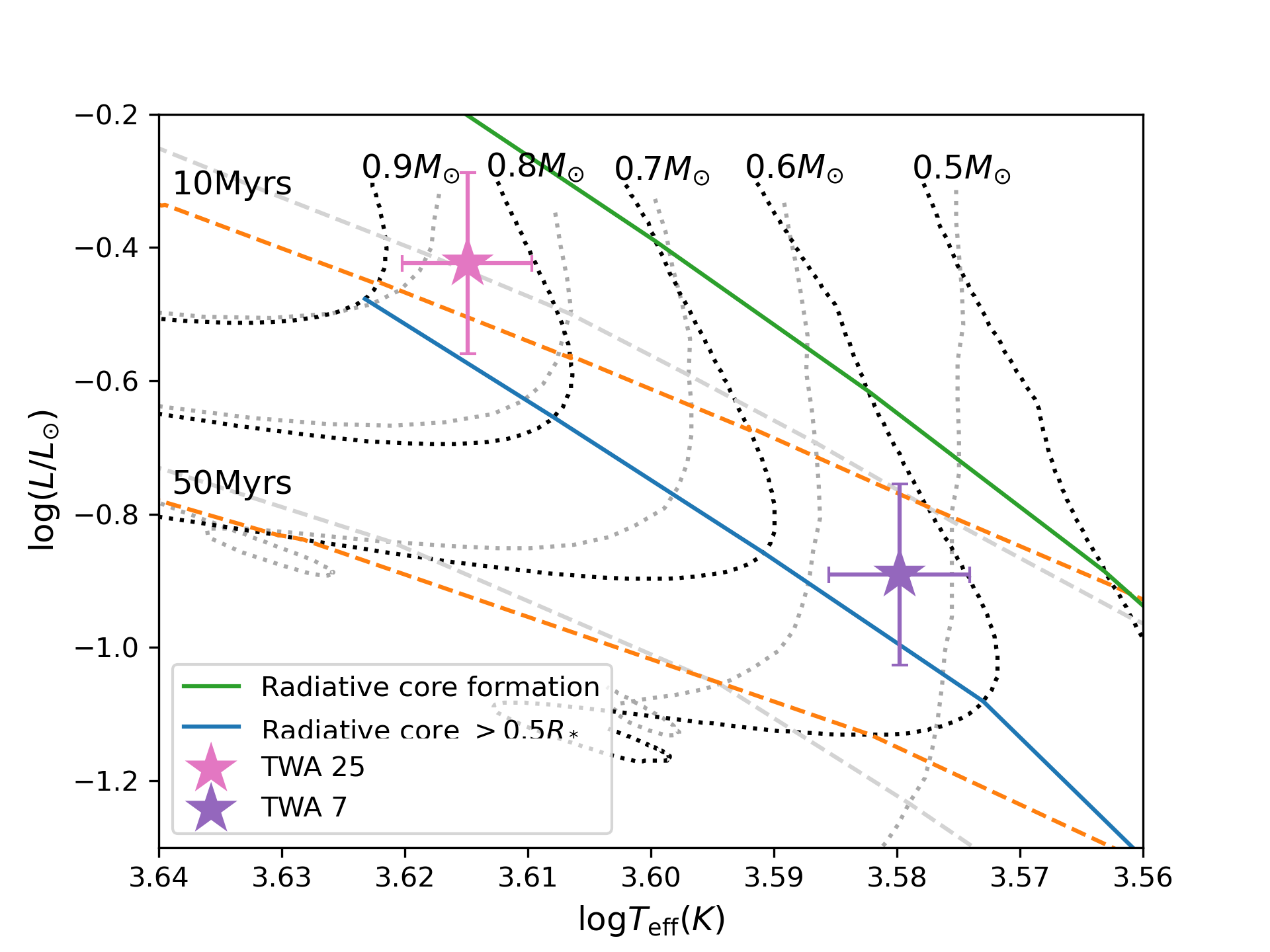}
 \caption{HR diagram showing the locations, including $1\sigma$ uncertainties, of TWA 25 (pink) and TWA 7 (purple) with relation to the \citet{Baraffe2015} PMS evolutionary models. The black dotted lines are evolutionary tracks for 0.9\Msun, 0.8\Msun, 0.7\Msun, 0.6\Msun and 0.5\Msun, the orange dashed lines are the 10Myr and 50Myr isochrones, the solid green line indicates the boundary between full convection and the formation of a radiative core, and the solid blue line shows the point at which that radiative core is greater than half the stellar radius. The grey dotted and dashes lines are the corresponding mass tracks and isochrones from the \citet{Siess2000} evolution tracks for comparison.}
 \label{fig:HRdiagramBoth}
 \end{figure}

Using our calculated luminosities and effective temperatures, we calculate mass, age and stellar radius for both stars based on the \cite{Baraffe2015} pre-main sequence stellar evolution models. The locations of TWA 25 and TWA 7 on the HR diagram with respect to these mass tracks and isochrones are shown in Figure \ref{fig:HRdiagramBoth}. Stellar masses, radii and ages derived for both stars are given in Table \ref{tab:stellar_param}.The age estimate for both stars are in agreement within 1 sigma with the average age of 10 Myr for the TW Hya Association \citep{Mentuch2008}. 

Investigating the internal structure of these stars, it can be seen in Figure \ref{fig:HRdiagramBoth} that these stars sit above the boundary where the radiative core grows to greater than half the stellar radius (solid blue line in), with TWA 25 estimated to have formed a radiative core of size $0.3\pm0.2$ R$_*$, and TWA 7 is expected to have a radiative core of $0.4\pm0.2$ R$_*$.

\subsection{Inclination}
\label{sec:inc}

For tomographic mapping, knowing the stellar inclination is crucial for interpreting how the variability in the LSD profiles relates to  the latitudes of recovered surface features. It is also one of the more challenging stellar parameters to determine. For TWA 25 and TWA 7, we consider both the estimates of debris disc inclination from imaging studies as a proxy for stellar inclination, and calculate the inclination based on stellar parameters. \cite{Choquet2016} imaged both TWA 25 and TWA 7 with NICMOS on the Hubble Space Telescope, and calculate a disc inclination of $75\pm6$ degrees for TWA 25 and $22\pm22$ degrees for TWA 7, where 0 degrees represents a face-on disc (and pole-on star if the disc and stellar rotation axes are aligned). TWA 7 has also been observed by \cite{Olofsson2018} who used SPHERE on ESO's Very Large Telescope, refining TWA 7's disc inclination to $13.1^{+3.1}_{-2.6}$ degrees.  

To test if these disc inclinations are reasonable approximations for the stellar inclination, we calculated the implied stellar radius given our measured  \vsini and \Prot values. For TWA 25 we find a theoretical radius of $1.23\pm0.05 R_{\odot}$, which agrees with our estimates from PMS evolution models within a $1\sigma$ uncertainty. For TWA 7, however we find a theoretical radius of $2.0^{+0.5}_{-0.4}$ \Rsun based on the Olofssson et al. inclination estimate. This is far larger than any radius suggested by the PMS models for a star of this temperature and luminosity, and is unlikely given the age of this star and star forming region. This is also true for radii calculated from inclinations $3\sigma$ above the disc inclination estimate. This suggests that the observed disc is misaligned with respect to the stellar rotation axis.

We next calculate the inclination from the measured \vsini\, and \Prot\, values in our data, and the stellar radius determined from our bolometric luminosities and the effective temperatures determined above. For TWA 25, we calculate an inclination of $81 \pm 63$ degrees, and for TWA 7 we get an inclination of $32\pm6$ degrees. Given the massive uncertainties, especially with the inclination of TWA 25, we leave inclination as a free parameter in our tomographic modelling of both stars for consistency, and it is these inclination values that are given in Table \ref{tab:stellar_param}.

 \section{Tomographic Modelling}
 \label{sec:TomMod}
 \subsection{Doppler Imaging}
 \label{sec:DI}

Using the Stokes I LSD profiles described in Section \ref{sec:LSD}, and the stellar parameters given in Table \ref{tab:stellar_param}, we reconstruct the surface brightness of TWA 7 and TWA 25 using the technique of Doppler Imaging (DI). For this we used the {\sc DOTS} code \citep{CollierCameron1997}, with modifications to reconstruct areas of both bright plage and cool spot, as described in \cite{Donati2014}. This code inverts the Stokes I data and applies a maximum entropy regularisation to determine the simplest map that best fits our data.  The surface brightness was reconstructed assuming a Milne-Eddington model atmosphere for the local line profile, and using mean Land\'e factor of 1.192, mean wavelength of 544nm (as set by the LSD profiles) and a linear limb darkening coefficient of 0.72. In this process we fit for the \vsini, line equivalent width (EW), radial velocity, rotation period and inclination for each star. The parameters are determined by pushing to a low value of reduced $\chi^2$ while varying the parameters, except for \vsini and EW, these are done separately. Radial velocity, period and inclination are fit first and then \vsini and EW are varied on a fine grid and noting the reduced $\chi^2$ values achieved within a set number of iterations (usually 20). The curves around our parameter grids show parabolic shapes for radial velocity, period and \vsini and EW, and so the parameters quoted and the associated uncertainties  are determined using parabolic fits to the chi-squared minimisation. For TWA 25 we estimate the average radial velocity, rotation period and \vsini to be $8.33\pm 0.25$ k\mps,  $5.07\pm0.03$ days and $11.9\pm 0.3$ k\mps, respectively. For TWA 7, we find an average radial velocity of $13.18\pm0.25$ k\mps, rotation period of $5.00\pm 0.01$ days and \vsini of $4.5\pm0.2$ k\mps. The inclination grids do not show a parabolic shape so this method could not be used to determine inclinations -  the plots show defined minima over a range of inclination angles, which are used as rough estimates of appropriate ranges of stellar inclination for these systems. For TWA 25 this is $57\pm5$ degrees, and $40^{+5}_{-15}$ degrees for TWA 7. We were unable to detect the presence of differential rotation in the Stokes I data of either star.

\begin{figure}
  \centering
    \includegraphics[width=.5\textwidth]{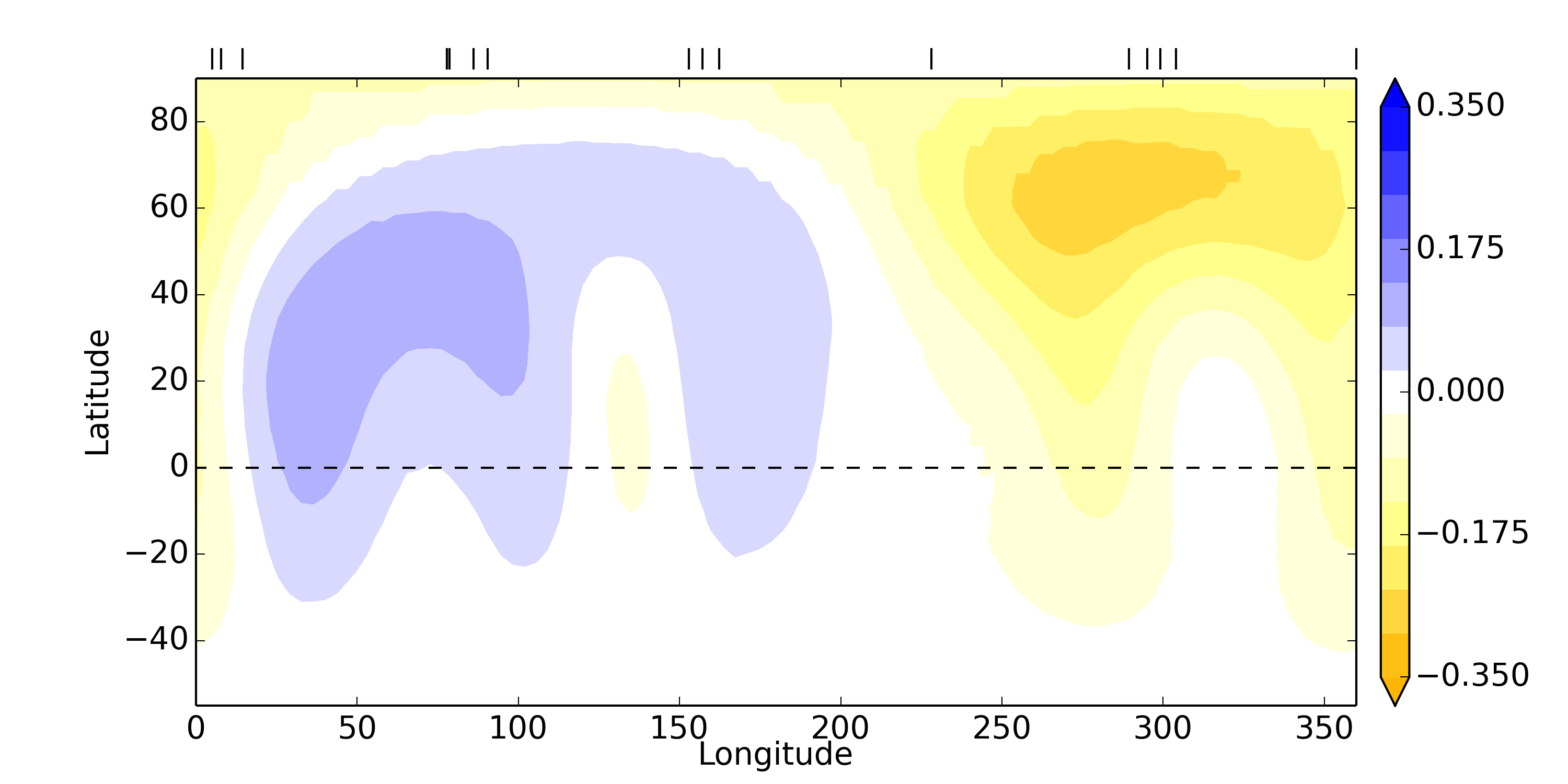}
  \caption{Surface brightness map of TWA 25, showing areas of bright plage (blue, $>0$), and areas of cool spot (yellow, $<0$), relative to the photosphere (0.0). The black dashed line indicates the equator, and the top tick marks indicate observed longitudes. }
  \label{fig:TWA25_final1_rec}
\end{figure}

\begin{figure}
  \centering
    \includegraphics[width=.5\textwidth]{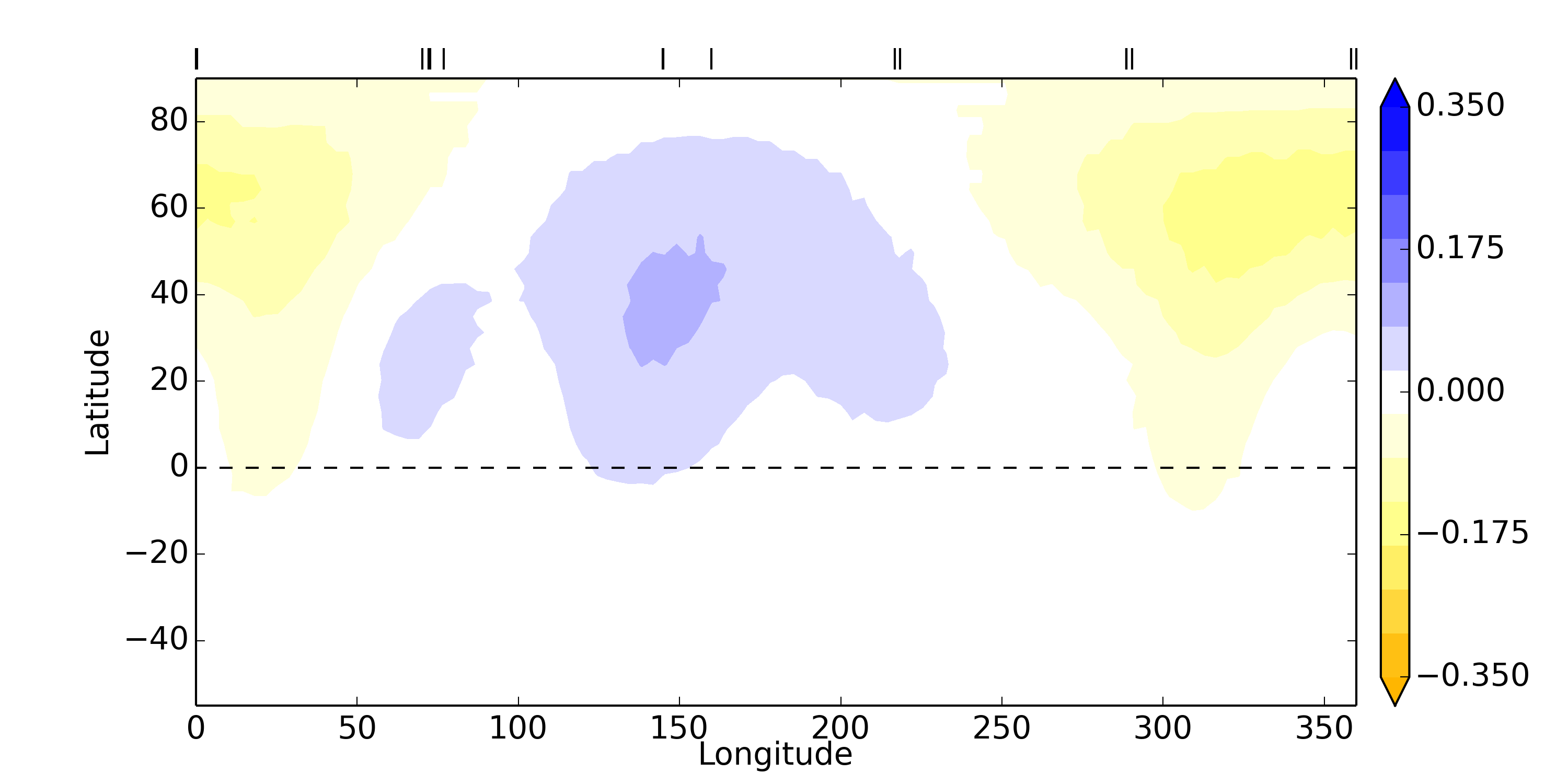}
  \caption{Surface brightness map of TWA 7, showing areas of bright plage (blue, $>0$), and areas of cool spot (yellow, $<0$), relative to the photosphere (0.0). The black dashed line indicates the equator, and the top tick marks indicate observed longitudes. }
  \label{fig:TWA7_final1_rec}
\end{figure}

Our resulting fits to the Stokes I profiles are shown in red in Figures \ref{fig:TWA25_final_fit} and \ref{fig:TWA7_final_fit}, and our reconstructed surface brightness maps are shown in Figures \ref{fig:TWA25_final1_rec} and \ref{fig:TWA7_final1_rec}. TWA 25 displays large areas of both bright plage and cool spots at high, mid and low latitudes. 

The reconstructed surface brightness maps of TWA 7 show a similar morphology, with one large region of cool spot, a large area of plage. The simplicity of the map is due to the lack of resolution of the stellar surface which results from the star having a low \vsini, which in turn limits the spatial resolution of the resulting map.

\subsection{Zeeman Doppler Imaging}
\label{sec:ZDI}

The large-scale surface magnetic fields are reconstructed for each star from their Stokes V LSD profiles using the Zeeman Doppler Imaging (ZDI) program {\sc ZDOTS} \citep{Hussain2000,Hussain2002,Hussain2016}, which is essentially DOTS applied to circularly polarised spectra. ZDOTS allows the surface magnetic field to be expressed as a series of spherical harmonics, and uses maximum entropy regularisation to find the simplest field configuration that best fits our Stokes V LSD profiles. The Stokes V profiles are modelled in a weak field approximation, using a mean Land\'e factor and central wavelength equivalent to those of the Stokes V LSD profiles. The stellar model, including line profile modelling and stellar parameters are identical in both DOTS and ZDOTS, and we take into account the surface brightness reconstructions produced above in our reconstruction of the stellar surface magnetic field maps. 

\begin{figure*}
  \centering
    \includegraphics[width=1\textwidth]{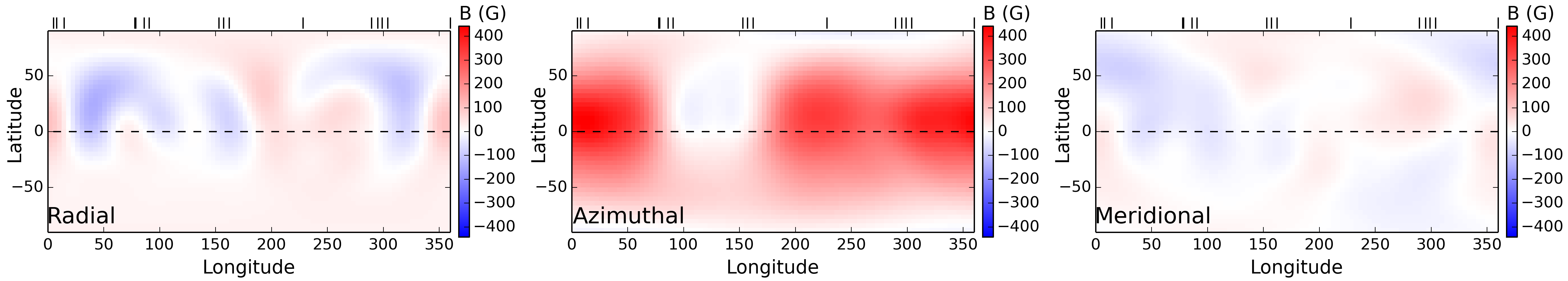}
  \caption{Large-scale radial (left), azimuthal (centre) and meridional (right) surface magnetic field maps of TWA 25 corresponding to the Stokes V fits in Figure \ref{fig:TWA_StokesV_fit_fin}. These results a show strong azimuthal field compared to the radial and meridional field components. The poloidal (radial and meridional) field  is simple and non-axisymmetric, with $80\%$ for the poloidal field energy contained in dipolar and quadrupolar components, and $\sim 97\%$ poloidal field energy in non-axisymmetric components. }
  \label{fig:TWA25_magmap_fin}
\end{figure*}

For TWA 25, the reconstructed radial, azimuthal and meridional components of the large-scale magnetic field are shown in Figure \ref{fig:TWA25_magmap_fin}, with the associated fit to the Stokes V profiles shown in Figure \ref{fig:TWA_StokesV_fit_fin}. Since it is easy of over-fit data in ZDI, an optimum reduced $\chi^2$ was determined using the method of \cite{AlvaradoGomez2015}, resulting in reduced $\chi^2$ of 1.124 for this fit. 

In the fitting process, we investigate the presence of differential rotation, and find a best fit equatorial rotation period of $5.01\pm 0.03$ days, and a shear value of $0.025 \pm 0.04 $ radians/day. Our reconstructed field has a mean total field strength of 535 G, and shows a far stronger azimuthal field compared to the radial or meridional fields, as well as a high level of non-axisymmetry in the radial component. We quantify this by examining the distribution of energy among the different spherical harmonic components. This solution indicates a dominantly toroidal field ($83\%$ energy in toroidal components). The poloidal field is highly non-axisymmetric, with $97\%$ of the poloidal field energy not aligned with the stellar rotation axis, but it is relatively simple, with $80\%$ of the poloidal field contained in the dipolar and quadrupolar components.

\begin{figure*}
  \centering
    \includegraphics[width=1\textwidth]{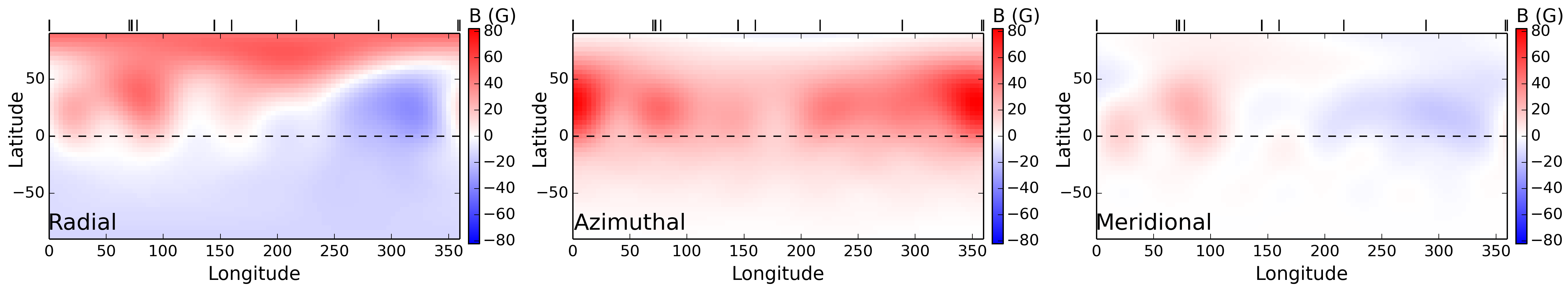}
  \caption{The large-scale radial (left), azimuthal (centre) and meridional (right) surface magnetic field maps of TWA 7 corresponding to the Stokes V  fits in Figure \ref{fig:TWA7_StokesV_fit_rvtweak2}. These results show a stronger azimuthal field than either the radial or meridional components. The poloidal (radial and meridional) field is quite simple with $90\%$ of the poloidal field energy in dipolar or quadrupolar components, and predominantly non-axisymmetry, with only $38\%$ of the poloidal field energy in axisymmetric components}
  \label{fig:TWA7_magmap_rvtweak}
\end{figure*}

The reconstructed large-scale surface magnetic field maps for TWA 7 are shown in Figure \ref{fig:TWA7_magmap_rvtweak}, and the fit to the Stokes V LSD profiles are shown in Figure \ref{fig:TWA7_StokesV_fit_rvtweak2}, fit to an optimum reduced $\chi^2$ of 1.21. Given the poorer signal-to-noise ratio (SNR) of these observations, we note in Figure \ref{fig:TWA7_StokesV_fit_rvtweak2} which profiles are definite magnetic detections, defined as having a false alarm probability (FAP) less than $10^{-5}$, a marginal detection with a FAP between $10^{-5}$ and $10^{-3}$, or a non-detection with FAP greater than $10^{-3}$. Not included in this figure are observations lacking the full set of 4 sub-exposures, as these are excluded from the ZDI analysis. 

We are unable to obtain information about any surface differential rotation, though we do obtain a best fit rotation period of $5.00\pm0.01$ days from a reconstruction using just the Stokes V profiles, which is in agreement with the rotation period determined using just the intensity profiles.  The resulting reconstructed large-scale magnetic field has a mean total field strength of 30 G, and, as with TWA 25, show a strong azimuthal band at mid latitudes. The meridional field is weaker, but this is typically the case with ZDI, due to some degree of cross-talk between the meridional and radial magnetic field components, though this effect is reduced by expressing the field as spherical harmonics \citep{Donati2001}.  We again quantify the large-scale by examining the percentage of energy divided among the different magnetic field components. The resulting field is similar to that of TWA 25, with a dominantly toroidal field ($34\%$ energy in poloidal components), and a very simple poloidal field with $90\%$ energy in dipolar and quadrupolar components. The poloidal field is also dominantly non-axisymmetric, with only $38\%$ for the poloidal field energy aligned with the rotation axis.

\section{Longitudinal Magnetic Field, and Activity indicators}
\label{sec:Blongs}
\begin{figure}
  \centering
    \includegraphics[width=.5\textwidth]{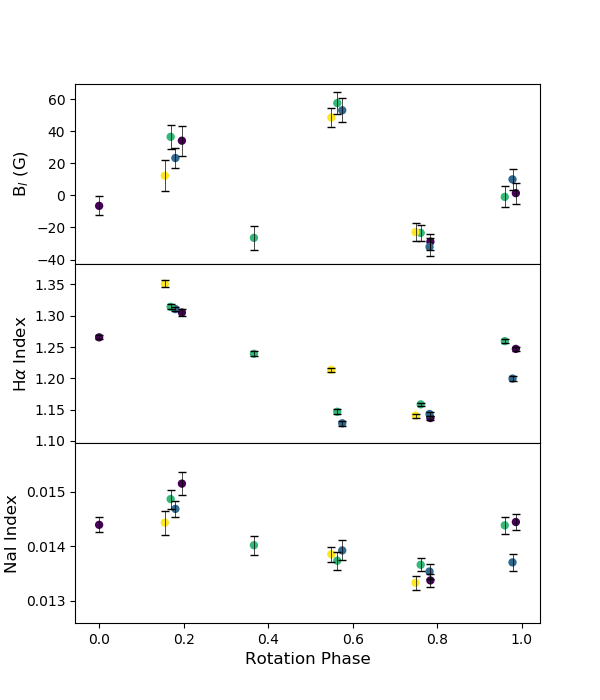}
  \caption{Plots of longitudinal magnetic field measurement (top), H$\alpha$ indices (middle) and Na I doublet indices (bottom) for TWA 25. The colours of the points represent the number of rotations, with purple being the first cycle, blue the second, green the third, and yellow the fourth observed cycle. }
  \label{fig:TWA25_Halpha_Blongs_Na_Phase}
\end{figure}

\begin{figure}
  \centering
    \includegraphics[width=.5\textwidth]{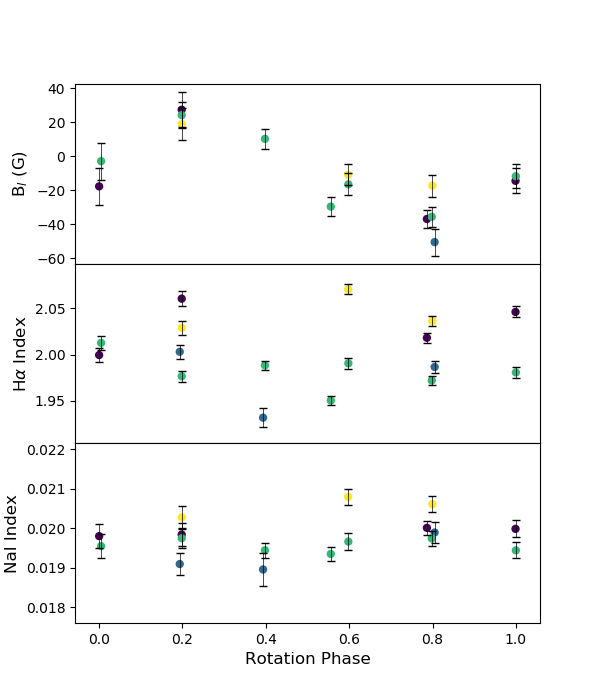}
  \caption{Plots of longitudinal magnetic field measurement (top), H$\alpha$ indices (middle) and Na I doublet indices (bottom) for TWA 7. Colours are as per Figure \ref{fig:TWA25_Halpha_Blongs_Na_Phase}}
  \label{fig:TWA7_Halpha_Blongs_Na_Phase}
\end{figure}

In addition to reconstructing the large-scale brightness and magnetic field morphologies, we explore other probes of stellar magnetic activity, namely the line-of-sight (longitudinal) magnetic field, $B_l$, and the H$\alpha$ and Na I doublet indices. These are plotted for both stars in Figures \ref{fig:TWA25_Halpha_Blongs_Na_Phase} and \ref{fig:TWA7_Halpha_Blongs_Na_Phase}. The longitudinal magnetic field is calculated for each Stokes V profile using the program of \cite{Grunhut2013}, which follows the method of \cite{Wade2000}, using the mean Land\'e factor and mean wavelength of our Stokes V LSD profiles. Uncertainties in $B_l$ are calculated by propagating the Stokes I and V LSD uncertainties. For TWA 7 we exclude incomplete observations (where only two of the four  sub-exposures in the sequence were obtained), as without a full set of sub-exposures we are unable to calculate the null polarisation profile needed to assess the level of spurious signal in our profiles. 

We define an H$\alpha$ index following \cite{Marsden2014}, using a rectangular emission bandpass of 0.36 nm, centred on 656.285 nm, and two continuum bandpasses in 0.22 nm, centred on 655.885 nm and 656.730 nm. For the Na I doublet indices we define the index as in \cite{GomesdaSilva2014}, with two rectangular emission bandpasses of width 0.1 nm centred on 589.592 nm and 588.995 nm, and two continuum bandpasses centred on 580.50 nm and 609.00 nm, with width 1.0 nm and 2.0 nm, respectively. All spectral indices are calculated in the stellar rest frame, with uncertainties calculated by propagation of the intensity spectra uncertainties. Figures \ref{fig:TWA25_Halpha_Blongs_Na_Phase} and \ref{fig:TWA7_Halpha_Blongs_Na_Phase} show $B_l$ values plotted with $1\sigma$ error bars, and H$\alpha$ and Na I doublet indices are shown with $3\sigma$ error bars for clarity. The colours of the points represent the number of rotations, with purple being the first cycle, blue the second, green the third, and yellow the fourth observed cycle. 

For both stars, we see a range of values in all activity measures, which is expected given the evolutionary states of these stars. TWA 25 shows variations in longitudinal magnetic field strengths, $|B_l|$, between between $1\pm7$ G and $58\pm7$ G. Significant variation is also observed in the H$\alpha$ indices, and to a lesser extent in the Na I doublet indices. Across all observations of TWA 25, $B_l$ measurements at a given observed phase have a small dispersion, indicating that the large-scale magnetic field is stable over the time span of our observations. This is also the case for the H$\alpha$ and Na I doublet indices, indicating that these changes are associated with particular features in the stellar surface, and that these features are also largely stable over the time span of our observations. 

TWA 7 also displays variation in the longitudinal magnetic field, with $|B_l|$ values ranging between  $3\pm11$ G and $50\pm8$ G. At a given observed phase, $B_l$ values are consistent, indicating that the large-scale field is stable over the timespan of our observations. The H$\alpha$ and Na I doublet indices, however, both display large dispersions at a given phase, with the first (purple points) and last (yellow points), showing systematically higher H$\alpha$ and Na I doublet index values than the other cycles. This indicates that the chromospheric active regions evolve more quickly than the large-scale photospheric magnetic field.

\begin{figure}
  \centering
    \includegraphics[width=.5\textwidth]{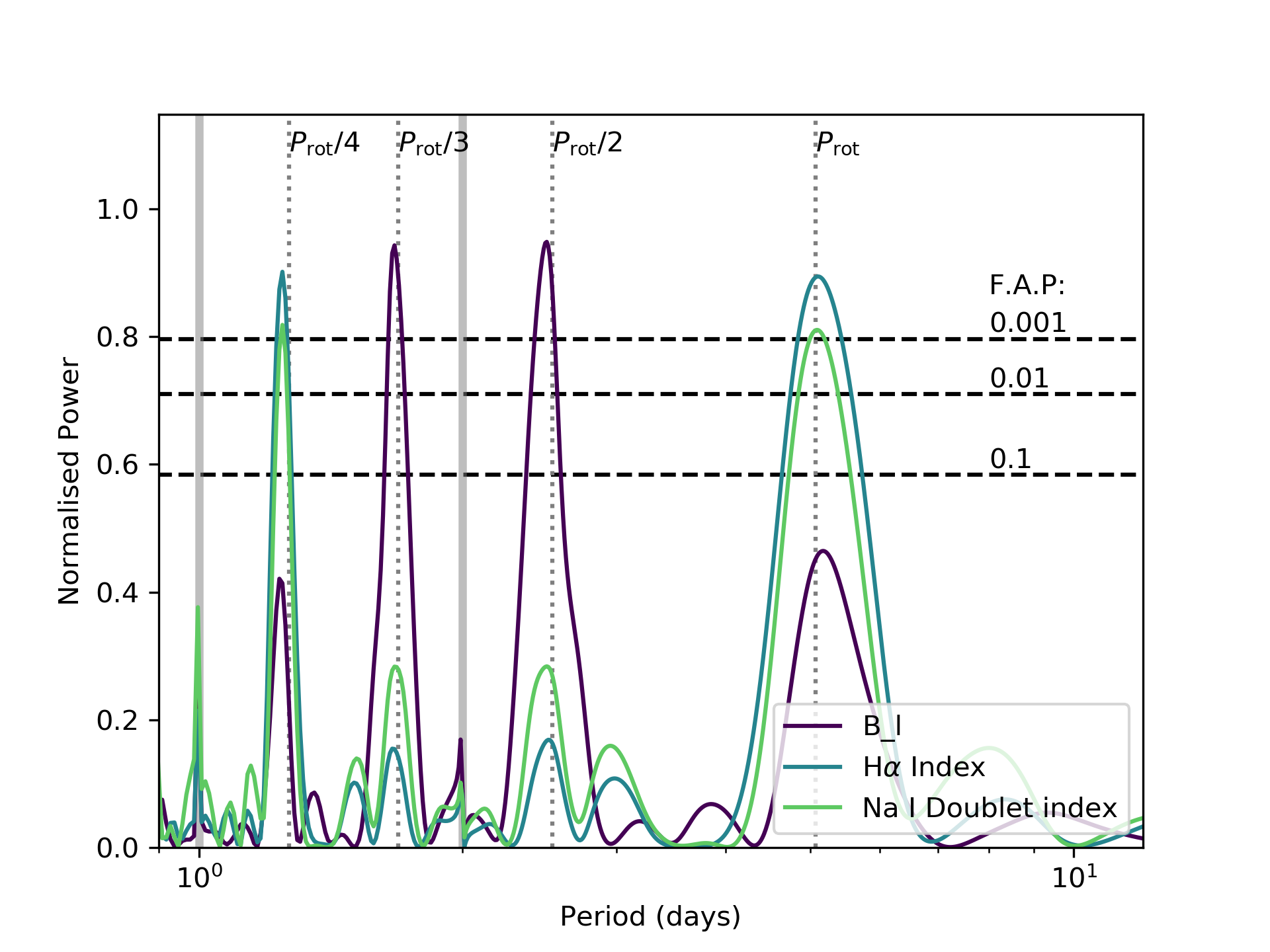}
  \caption{A generalised Lomb-Scargle periodogram of $B_l$, H$\alpha$ indices and Na I Doublet indices for TWA 25.  }
  \label{fig:Periodogram_TWA25_BlHaNa}
\end{figure}

\begin{figure}
  \centering
    \includegraphics[width=.5\textwidth]{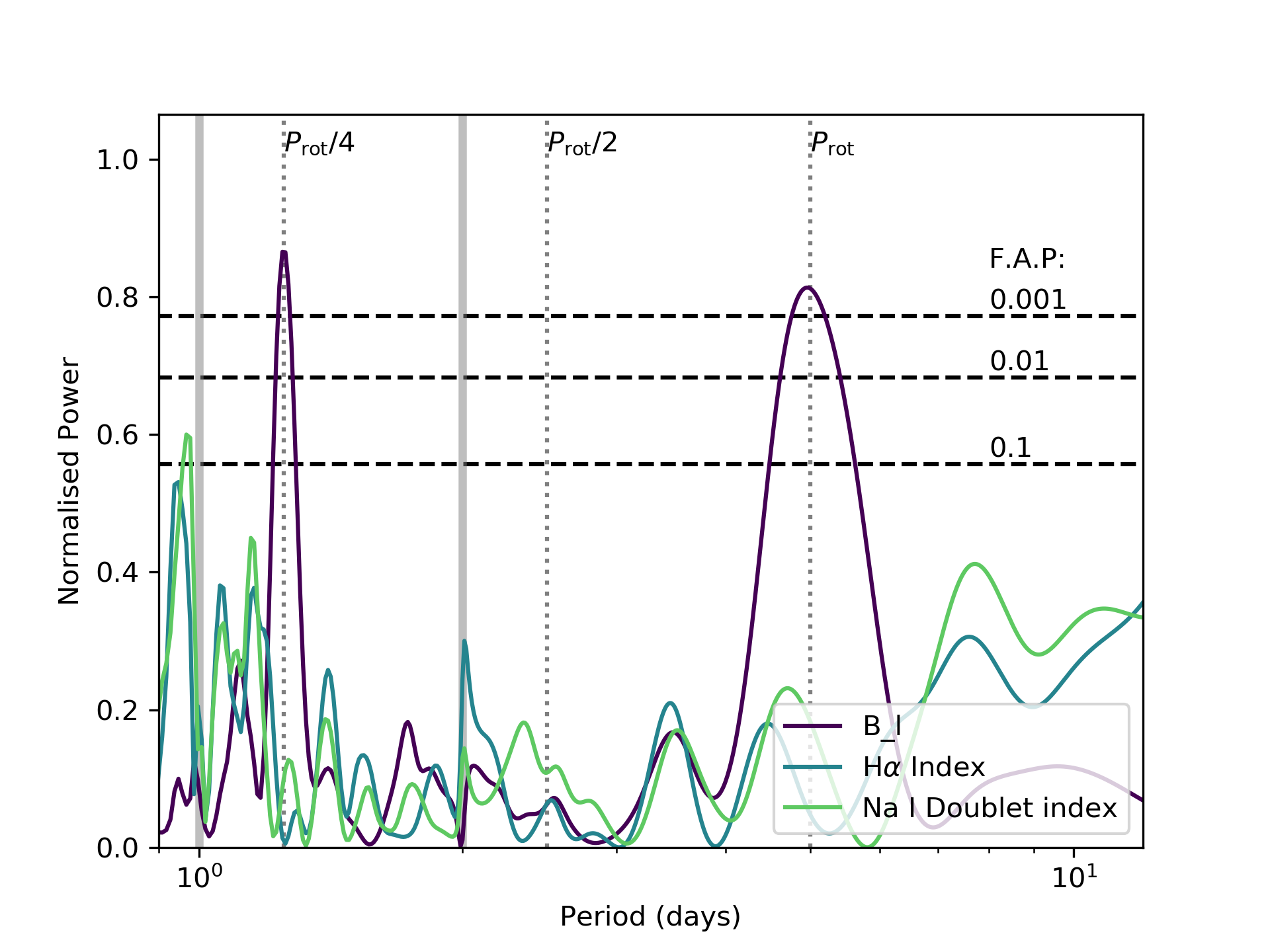}
  \caption{A generalised Lomb-Scargle periodogram of $B_l$, H$\alpha$ indices and Na I Doublet indices for TWA 7.  }
  \label{fig:Periodogram_TWA7_BlHaNa}
\end{figure}

We analyse the periodicity of our $B_l$, H$\alpha$ and Na I doublet indices for both stars with a Generalised Lomb Scargle periodogram \citep{Zechmeister2009}, using the PyAstronomy\footnote{\url{https://github.com/sczesla/PyAstronomy}} package. These are shown in Figure \ref{fig:Periodogram_TWA25_BlHaNa} for TWA 25 and in Figure \ref{fig:Periodogram_TWA7_BlHaNa} for TWA 7. Included in these figures are vertical dotted lines indicating the stellar rotation period and fractions of the rotation period, as well as horizontal dashed lines indicating the false alarm probability (FAP) levels. These FAP values are calculated in the PyAstronomy program as per Equation 24 of \citet{Zechmeister2009}. Both data sets are sampled approximately daily, giving an associated Nyquist frequency of 0.5 d$^{-1}$. These are highlighted as grey lines at periods of 1 and 2 days, respectively.  

The periodograms of $B_l$, H$\alpha$ and Na I doublet indices for TWA 25 all show peaks around the stellar rotational period. For the H$\alpha$ and the Na I doublet, this is the second highest peak (the highest is around the 1/4 rotation period), but is still detectable above a FAP of 0.001. For the $B_l$ values however, the peak at the rotation period is small. Instead, the highest peak in $B_l$ is at 1/2 the rotation period, with the second highest peak at 1/3 the rotation period. The dominance of these peaks at fractional aliases of the full rotation period reflects the magnetic field geometry: there are multiple magnetic regions that come into and out of the line of sight in a given rotation, giving signal power at fractions of that rotation period. All periodograms show a spike around the 1 day sampling rate, and small peaks at the 2 day period associated with the Nyquist frequency. 

For TWA 7, the periodogram of $B_l$ has the highest peak at 1/4 of the stellar rotation period with detectability at the FAP level of 0.001, and the second highest peak at the stellar rotation period. {This is reflective of the dominantly dipolar and quadupolar field reconstructed, which shows multiple, regularly spaced magnetic regions, adding power at this fraction of the total rotation period.} The H$\alpha$ and Na I doublet indices, however, do not have any peaks of significant power associated with the stellar rotation period. Instead, their highest peaks are around the cadence of observations of 1 day. 

Taking the periodograms of both stars together shows that, while activity indices like Na I and H$\alpha$ can be useful in determining the stellar rotation period (e.g. TWA 25), this is not always the case (e.g. TWA 7), as as also found by \cite{Hebrard2016}. Thus, monitoring of the longitudinal magnetic field is an effective method of measuring stellar rotation, independent of the behaviour of the chromospheric activity indices.

\section{Radial velocities}
\label{sec:radvel}
\begin{figure}
  \centering
    \includegraphics[width=.5\textwidth]{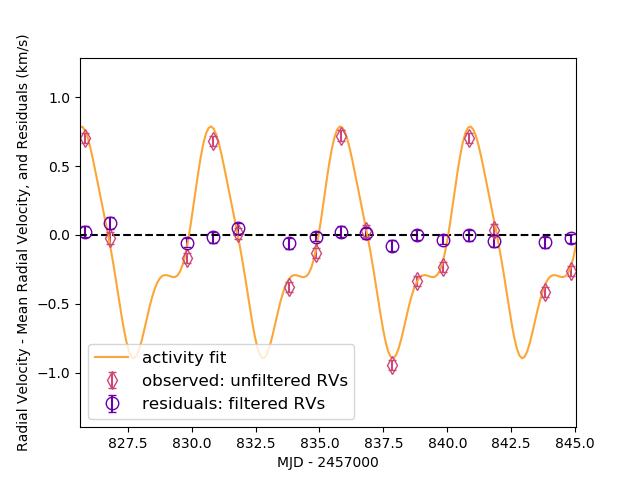}
  \caption{Radial velocities of TWA 25. The pink diamonds are the measured radial velocities of the LSD profiles (black lines in Figure \ref{fig:TWA25_final_fit}), the orange line is the radial velocity contribution from the surface brightness inhomogeneities, as measured from our fit to the surface brightness (red lines in Figure \ref{fig:TWA25_final_fit}), and the purple circles are the activity-filtered radial velocities, calculated by subtracting the brightness fit (orange line) from the measured radial velocity (pink diamonds). The RMS in the raw radial velocities is 465 \mps, which is further reduced to 42 \mps after filtering, of the same order as the mean uncertainty of 38 \mps. }
  \label{fig:Radvel_TWA25_final_line}
\end{figure}

\begin{figure}
  \centering
    \includegraphics[width=.5\textwidth]{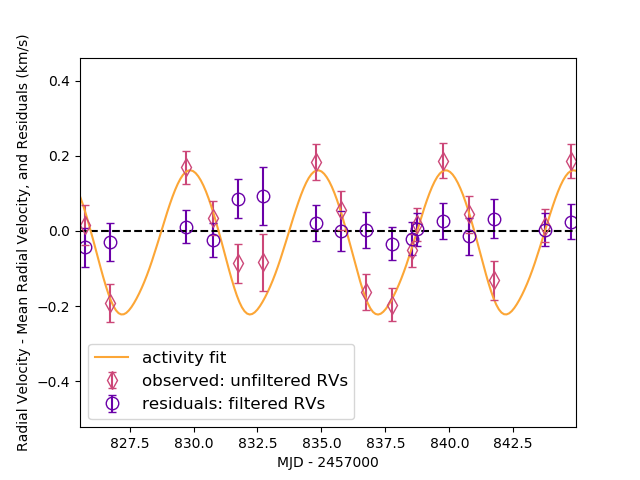}
  \caption{Radial velocities of TWA 7. The pink diamonds are the measured radial velocities of the LSD profiles (black lines in Figure \ref{fig:TWA7_final_fit}), the orange line is the radial velocity contribution from the surface brightness inhomogeneities, as measured from our fits based on our surface brightness reconstruction (red lines in Figure \ref{fig:TWA7_final_fit}), and the purple circles are the activity filtered radial velocities, calculated by subtracting the brightness fit (orange line) from the measured radial velocity (pink diamonds). The RMS in the raw radial velocities is 127 \mps, which is reduced to 36 \mps after filtering, i.e. smaller than the mean error of 49 \mps. }
  \label{fig:Radvel_TWA7_final_line}
\end{figure}

We analyse the radial velocities of TWA 25 by measuring the first order moment of each Stokes I profiles in the heliocentric rest frame, with uncertainties calculated as in \cite{Butler1996}. These RV values are shown as pink diamonds in Figure \ref{fig:Radvel_TWA25_final_line}. We then use the fits from our surface brightness maps to filter out the activity-induced RV signal, by measuring the RVs of our DI fits (shown as an orange line in Figure \ref{fig:Radvel_TWA25_final_line}), and subtracting them from the radial velocities of our observations, as per the method of \cite{Donati2015}. The resulting residuals are shown as purple circles in Figure \ref{fig:Radvel_TWA25_final_line}. In doing this, we reduce the RMS of the radial velocity variations from 465 \mps to 42 \mps, which is of the same order as the mean uncertainty of 38 \mps. The semi-amplitude of the radial velocities is reduced from 833 \mps to 81 \mps.  The same radial velocity analysis was performed for TWA 7, and this is plotted in Figure \ref{fig:Radvel_TWA7_final_line}. The semi-amplitude of TWA7's radial velocities decreased from 191 \mps to 68 \mps, and the RMS of the radial velocities was reduced from 127 \mps to 36 \mps, below than the average uncertainty of 49 \mps. 


\begin{figure}
  \centering
    \includegraphics[width=.5\textwidth]{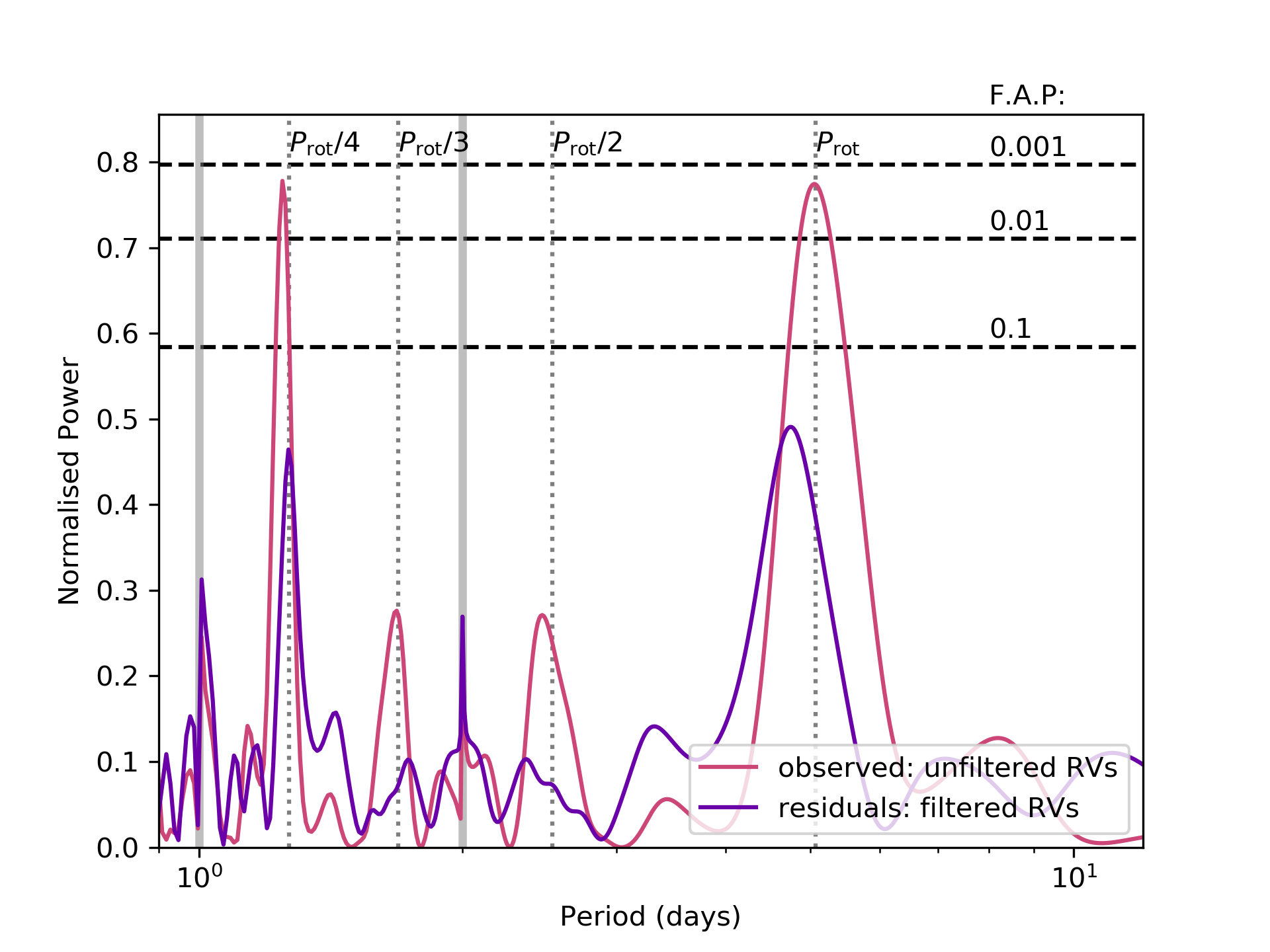}
  \caption{A generalised Lomb-Scargle periodogram of the radial velocities of TWA 25, and residuals after filtering for activity. }
  \label{fig:Periodogram_TWA25_Radvels}
\end{figure}

\begin{figure}
  \centering
    \includegraphics[width=.5\textwidth]{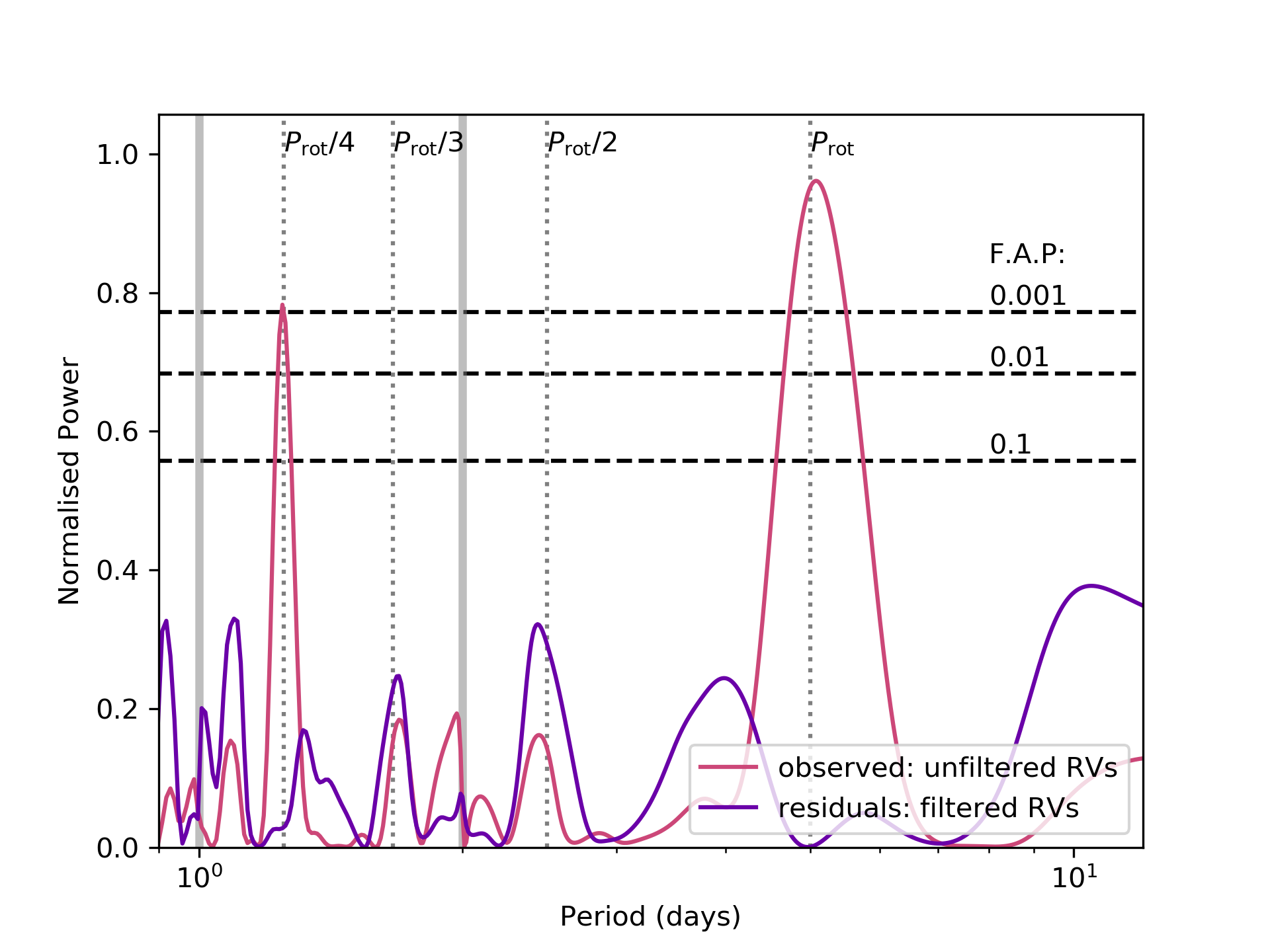}
  \caption{A generalised Lomb-Scargle periodogram of the radial velocities of TWA 7, and residuals after filtering for activity.  }
  \label{fig:Periodogram_TWA7_Radvels}
\end{figure}

The periodicity of the radial velocities and activity filtered residuals were analysed for both stars with a Generalised Lomb Scargle periodogram using the same procedure as used for the activity indices. These are shown in Figures \ref{fig:Periodogram_TWA25_Radvels} and \ref{fig:Periodogram_TWA7_Radvels} for TWA 25 and TWA 7, respectively. The radial velocity periodograms for both stars show significant peaks at their rotation periods and the quarter harmonic of that period. The periodograms of the residual, activity filtered radial {velocities} have reduced power at those periods in both cases, but most notably in TWA 7, despite it being a challenging Doppler imaging target with a low \vsini\ and poorer signal-to-noise.

 \begin{figure}
  \centering
    \includegraphics[width=\columnwidth]{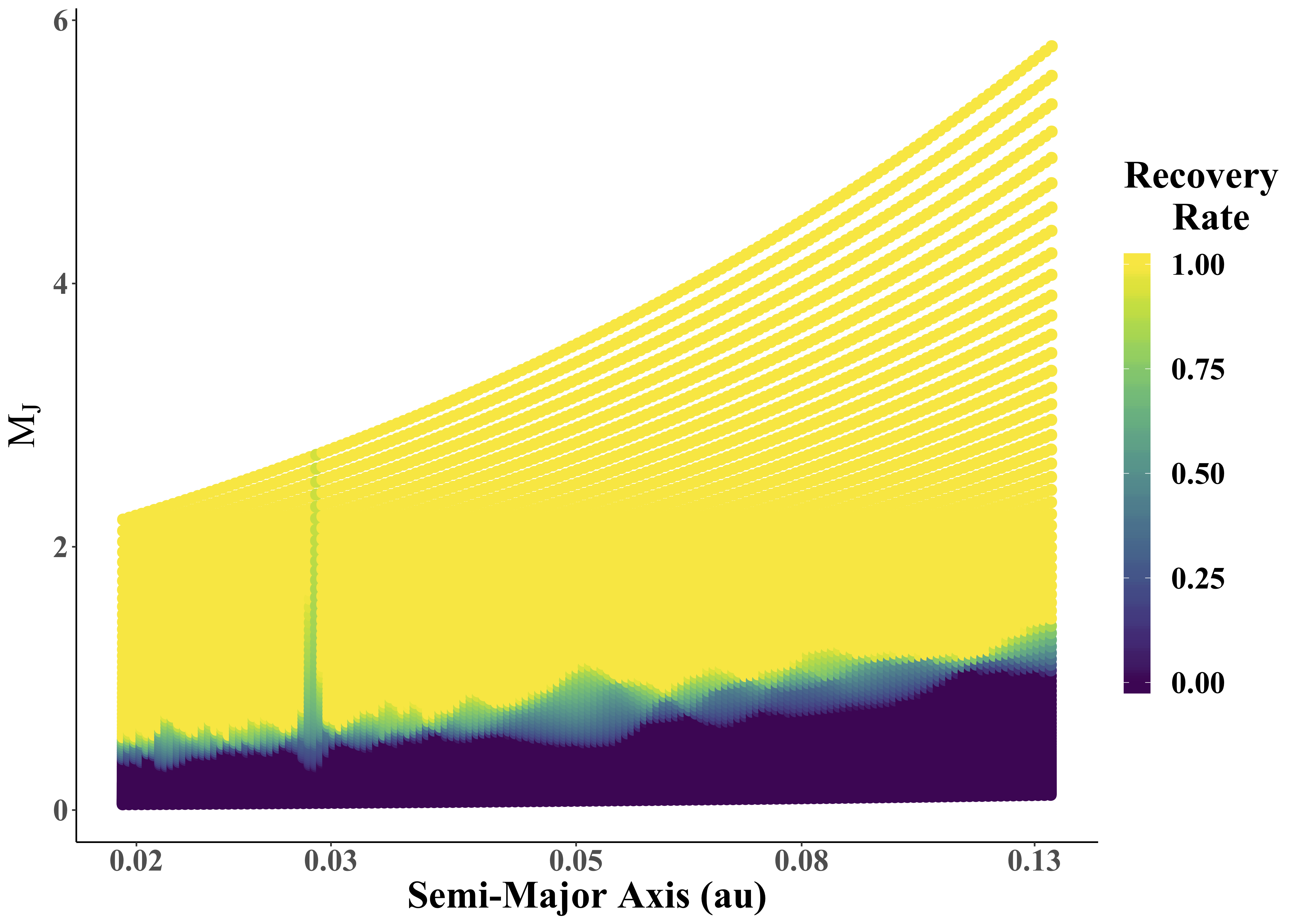}
    \includegraphics[width=\columnwidth]{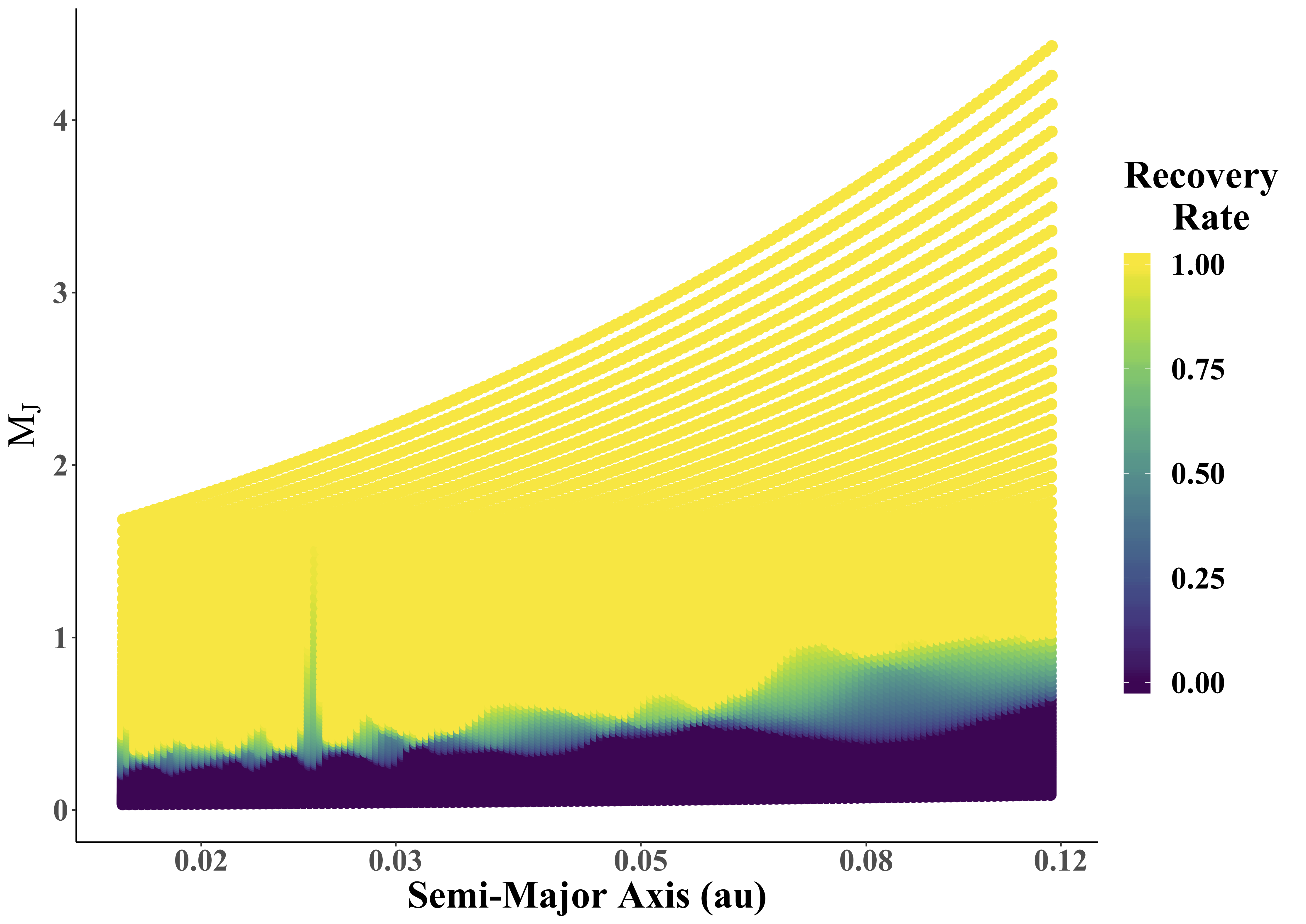}
  \caption{Detection probabilities for exoplanets orbiting TWA 25 (top panel), and TWA 7 (bottom panel), derived from the filtered RV data. }
  \label{fig:limitsplots}
\end{figure}

Planet detection limits were determined for both stars based on the activity-filtered radial velocities by adding an artificial Keplerian signal to each data set, then performing an F-test to ask whether the two data sets (original and with added planetary signal) are significantly different at a 99\% confidence level.  This approach has been used in cases where we have very limited data (usually $N<40$) and the traditional GLS periodogram recovery method becomes unreliable \citep{ppps8}.  This injection-recovery method is otherwise identical to that used in related exoplanet detection-efficiency studies \citep[e.g.][]{limitspaper,newjupiters,debris}.  Given that the velocity data for both stars span only 19 days, we test orbital periods $P<20$ days, corresponding to a maximum semimajor axis of $a\sim 0.13$ au.  We further assume that close in planets would arrive there by disc migration, and so we test for circular orbits only.  For each of the 20 trial orbital periods and 100 velocity semi-amplitudes $K$, we injected a circular Keplerian signal.  For each $(P,K)$ pair, we tested 1000 values of the orbital phase and performed an F-test comparing the original data with that to which the signal was added.  The result is a recovery fraction for each $(P,K)$ pair, as shown in Figure~\ref{fig:limitsplots}. The spike seen in each data set occurs at an orbital period of 2 days, and so results from the sampling of our data and its associated Nyquist frequency of 0.5 d$^{-1}$. It should be stressed that since we perform these tests on the activity filtered data, we are insensitive to any signal around the rotation period of each star, as signals around those periods are removed through the filtering process. Aside form planets on those periods, in general, we can rule out planets at the 99\% confidence level with M$_p \sin i\gtsimeq1.0$ \Mjup\ in close orbits around TWA 25, and M$_p \sin i\gtsimeq0.6$ \Mjup\ for TWA 7.  That is, our activity-corrected data permit the secure detection of close-in  planets with $K\sim$100 \mps for either star. 

\section{Summary and Discussion}
\label{sec:summary}

This paper presents the detailed analysis of high resolution spectropolarimetric data of 2 M-type wTTS, TWA 25 and TWA 7. Observations were taken between 12 to 31 March 2017, with a total of 16 observations for TWA 25, and 17 observations for TWA 7. Using Doppler imaging techniques, we reconstruct the large-scale surface brightness distributions for TWA 25 and TWA 7, and the large-scale magnetic fields for TWA 25 and TWA 7. In addition to this, we probe the stellar magnetic activity as a function of stellar rotation, by measuring the line-of-sight surface magnetic field, H $\alpha$ emission and core emission in the Na I doublet. We examine the radial velocity variations from both targets, and filter the stellar activity jitter from the radial velocity curves using our reconstructed brightness maps.

\subsection{Dynamos in PMS stars}

\begin{figure*}
  \centering
    \includegraphics[width=1.0\textwidth]{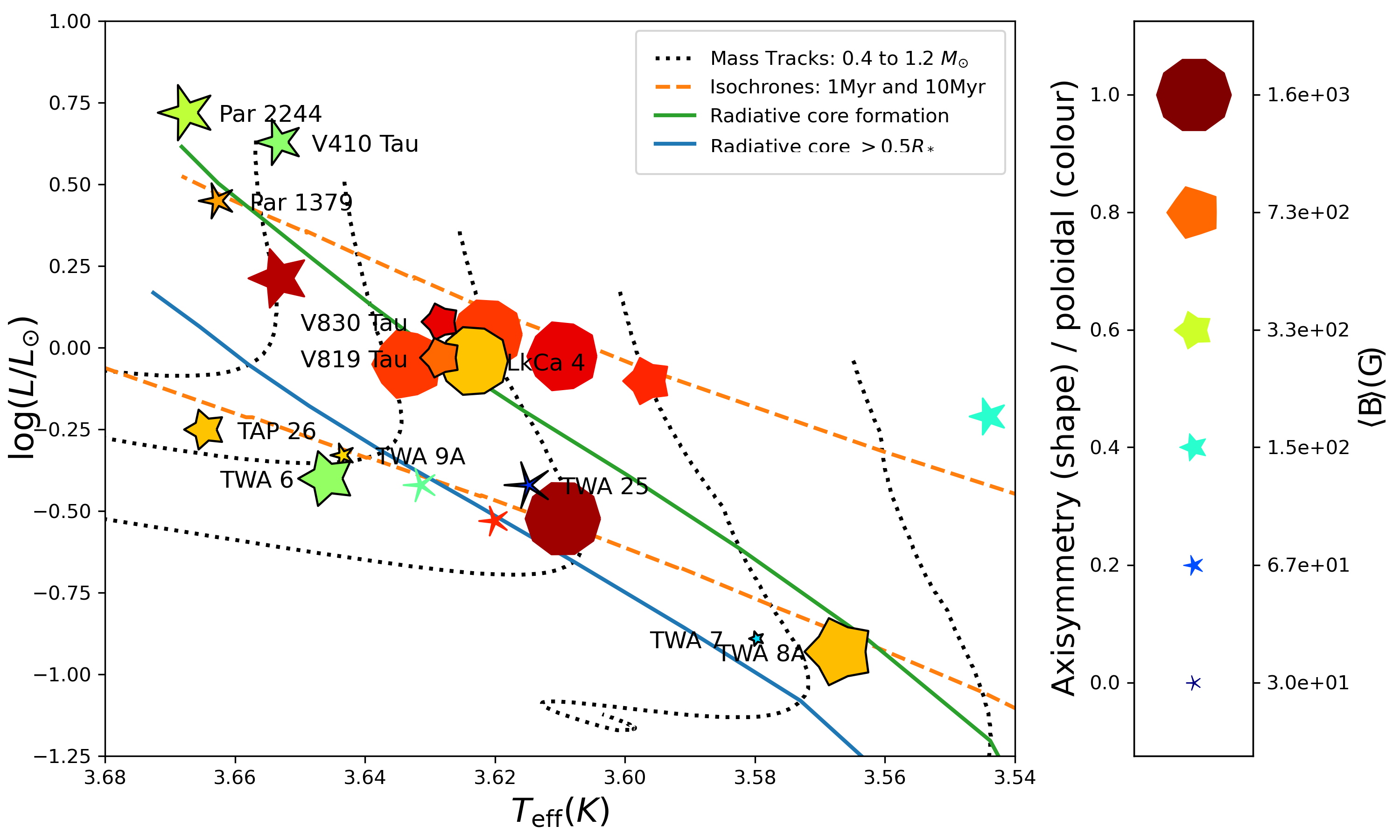}
  \caption{A Hertzsprung-Russell Diagram showing the updated MaTYSSE sample of WTTSs \citep[outlined in black; additional data from][]{Donati2014, Donati2015, Yu2017, Hill2017, Nicholson2018,Hill2019,Yu2019}, and the MaPP sample of CTTSs \citep[data from][]{Donati2007,Donati2008,Donati2010,Donati2010b,Donati2011,Donati2011b, Donati2012, Donati2013}. The size of each point represents the mean magnetic field strength, the colour indicates the percentage of magnetic field energy in poloidal components, and the shape indicates the percentage of poloidal field energy in axisymmetric components. The point for V410 Tau uses the mean of the 6 observations made by \citet{Yu2019}. The black dotted lines show the (right to left) 0.4, 0.6, 0.8, 1.0 and 1.2 \Msun mass tracks from the \citet{Baraffe2015} PMS evolution models, as well as 1 Myr (top) and 10Myr (bottom) isochrones as orange dashed lines. The boundary in evolution where a radiative core begins to form in indicated by the solid green line, and where the radiative core becomes greater than half the radius of the star is given by the solid blue line.  }
  \label{fig:Confusagram_Baraffe}
\end{figure*}

The magnetic field reconstructions for TWA 25 and TWA 7 add to the diversity of surface magnetic field morphologies found from spectropolarimetric studies of wTTS and in the wider PMS population. The strong toroidal fields reconstructed in TWA 25 and TWA 7 are similar to those seen in other wTTSs but differ significantly from CTTS with similar stellar properties. TWA 25 and TWA 7 are plotted in Figure \ref{fig:Confusagram_Baraffe} alongside the rest of the WTTS  (labelled and outlined in black), and the cTTS. WTTS display a larger range  of magnetic field morphologies compared to cTTSs. While the large scale fields of cTTS appear to show a dependence on the internal structure, the magnetic fields of wTTS display no such trends \citep{Gregory2012}. 

These  differences in the observed magnetic field behaviour of wTTS and cTTS suggest potential differences in the dynamos of accreting and non-accreting stars. It is interesting to note, however, that the dynamo models of \cite{EmeriauViard2017} do not include accretion processes in their simulations, and yet produce results that are only consistent for the observed sample of accreting stars.  Given wTTS overlap with the MaPP cTTSs in terms of their ages and stellar parameters, this sample is effectively also biased towards targets that have cleared their discs within approximately 10Myr, and have like different initial conditions of formation. Simulations of magnetic field generation in low mass stars report a `bistability' in the dynamo simulations \citep{Simitev2009}, where the resulting field configuration (either strong, poloidal and axisymmetric, or weak, toroidal, and non-axisymmetric) is sensitive to initial conditions, and a similar phenomenon may also explain the behaviour observed here.

The wider range in morphologies among the wTTSs compared to cTTSs could be explained by their wider range in rotation period.  The rotation rates among cTTS are more homogeneous and are on average slower, as the coupling to the disc and processes that dissipate the angular momentum of the star mean that they do not reach the same rotation speeds as the wTTS that are not inhibited by these mechanisms as they contract and spin up during transition towards the main sequence. There is a tentative relation among the classical T Tauri stars between rotation and morphology: the slower rotating cTTS are more likely to have large symmetric poloidal fields.  However, we see wTTSs with similar rotation periods and vastly different magnetic field topologies, such as TWA 8A, TWA 7 and TWA 25 that all have rotation periods around $\sim5$ days, but have very different large-scale magnetic fields. TWA 8A has a large, dominantly poloidal and axisymmetric field, where as TWA 7 has a small, dominantly toroidal and non-axisymmetric field, and TWA 25 has the most Toroidal field of the T Tauri star sample. This suggests we cannot explain the variety in the scale fields in wTTS by the variety of their rotation periods. 

Another possible explanation for the wide range in morphologies among the wTTS population is temporal variability. For the two stars presented in this paper, and recently all the other stars in the MaTYSSE sample are snapshot observations of these stars. \cite{Yu2019} have since published a series of 6 spectropolarimetric observations of V410 Tau between December 2008 and January 2016\footnote{An animated GIF depicting the variation of V410 Tau in the context of the updated TTS HR diagram can be found here: \url{https://tinyurl.com/wcdjseps}}. Over these observations, V410 Tau varies in its degree of poloidal and toroidal field, and in the degree of axisymmetry, similar to but not as wide as the range of values observed across the whole wTTS population; TWA 25 is still the most toroidal least axisymmetric observed in the MaTYSSE sample. The observed mean magnetic field strength changes only marginally over the observations of V410 Tau, indicating that the observed differences in field strengths across the MaTYSSE wTTS sample are due to other phenomena. 

We are unable to draw any definitive conclusions here, as the observed sample of T Tauri stars is still quite small. However, the advent of new infrared high-resolution spectropolarimeters, such as SpIRou on CFHT \citep{Donati2020} and CRIRES+ on the ESO VLT \citep{Follert2014}, presents exciting opportunities in the study of T Tauri star magnetic fields, and particularly for the  M-type stars in this class. These new instruments will provide useful insights into these cooler stars in greater numbers than is possible at optical wavelengths, allowing us to build statistically significant samples and also to further explore temporal variability in the large-scale magnetic fields of PMS stars.

\subsection{Radial velocities and stellar activity}

TWA 25 and TWA 7 both exhibit large, periodic variations in their radial velocities. In both stars, the surface spot and plage regions create large distortions in the spectral absorption line, which we are able to characterise with DI, and filter their contribution to the RV variability. For TWA 25, we reduce the RMS of our RVs from 465 \mps to 42 \mps, which is in line with our mean uncertainty of 38 \mps, and for TWA 7 reduce the RMS from 127 \mps to 36 \mps, less than the average uncertainty of 49 \mps. Given the amplitude of these filtered radial velocities, the period and sampling of observations, and the uncertainties on our radial velocity measurements, for TWA 25 we can rule out a planet with M$_p \sin i$ greater than 1.0 Jupiter masses, orbiting closer than 0.13 au, and for TWA 7 we can rule out out a planet with M$_p \sin i$ greater than 0.6 Jupiter masses, orbiting closer than 0.13 au.  We thus exclude the presence of close in giant planets for both of these stars. 

TWA 7 is an example of a `false positive' planet signal in the radial velocities of an active star. It shows large, sinusoidal radial velocity variations without obvious line distortion, as is typical for stars at these inclinations.  At face value it is easy to assume that such a motion could be caused by a planet, and the lack of periodicity in activity indices and small amplitudes in line bisectors compared with the RV amplitude supports this. However, given the variability in our circularly polarised spectra, which trace the variations in longitudinal magnetic field, we can infer that this radial velocity signal is likely caused by stellar activity. It is reminiscent of the case of TWA Hya, where a hot Jupiter was hypothesised to be orbiting on the same period as the stellar rotation period \citep{setiawan2008}, but this signal was later shown to be caused by a dark surface feature \citep{Donati2011}. This study further highlights the usefulness of longitudinal magnetic field measurements for indicating the stellar activity origins of radial velocity variations, where traditional chromospheric activity indices might fail. 

Given the lack of exoplanet detections for both stars, we can adjust the observed occurrence rate of hot Jupiters around wTTSs from 1 in 5 to 1 in 6, as measured by the MaTYSSE sample (2 detections within the now total sample of 12). This is still dramatically higher than the MS occurrence rate of 1\%, but the MaTYSSE sample is still very small, and a larger sample is needed to estimate the true occurrence rate. The new generation of high resolution spectropolarimeters mentioned above will help greatly with establishing more robust statistics of hot Neptunes and Jupiters around young stars.

\section*{Acknowledgements}
{Thanks to Louise Yu for her help with the stellar luminosities. }This work has made use of the VALD database, operated at Uppsala University, the Institute of Astronomy RAS in Moscow, and the University of Vienna. This work has made use of data from the European Space Agency (ESA) mission {\it Gaia} (\url{http://www.cosmos.esa.int/gaia}), processed by the {\it Gaia} Data Processing and Analysis Consortium (DPAC, \url{http://www.cosmos.esa.int/web/gaia/dpac/consortium}). Funding for the DPAC has been provided by national institutions, in particular the institutions participating in the {\it Gaia} Multilateral Agreement. We also warmly thank the IDEX initiative at Universit\'e F\'ed\'erale Toulouse Midi-Pyr\'en\'ees (UFTMiP) for funding the STEPS collaboration programme between IRAP/OMP and ESO and for allocating a ‘Chaire d'Attractivit\'e’ to GAJH, allowing her to regularly visit Toulouse to work on MaTYSSE data. This research is supported by USQ's Strategic Research Initiative programme. 

\section*{Data Availability}
The data underlying this article can be accessed from the European Southern Observatory Science Archive. The derived data generated in this research will be shared on reasonable request to the corresponding author.




\bibliographystyle{mnras}
\bibliography{Mtype_wTTSs_bib} 

\begin{thebibliography}{}
\makeatletter
\relax
\def\mn@urlcharsother{\let\do\@makeother \do\$\do\&\do\#\do\^\do\_\do\%\do\~}
\def\mn@doi{\begingroup\mn@urlcharsother \@ifnextchar [ {\mn@doi@}
  {\mn@doi@[]}}
\def\mn@doi@[#1]#2{\def\@tempa{#1}\ifx\@tempa\@empty \href
  {http://dx.doi.org/#2} {doi:#2}\else \href {http://dx.doi.org/#2} {#1}\fi
  \endgroup}
\def\mn@eprint#1#2{\mn@eprint@#1:#2::\@nil}
\def\mn@eprint@arXiv#1{\href {http://arxiv.org/abs/#1} {{\tt arXiv:#1}}}
\def\mn@eprint@dblp#1{\href {http://dblp.uni-trier.de/rec/bibtex/#1.xml}
  {dblp:#1}}
\def\mn@eprint@#1:#2:#3:#4\@nil{\def\@tempa {#1}\def\@tempb {#2}\def\@tempc
  {#3}\ifx \@tempc \@empty \let \@tempc \@tempb \let \@tempb \@tempa \fi \ifx
  \@tempb \@empty \def\@tempb {arXiv}\fi \@ifundefined
  {mn@eprint@\@tempb}{\@tempb:\@tempc}{\expandafter \expandafter \csname
  mn@eprint@\@tempb\endcsname \expandafter{\@tempc}}}

\bibitem[\protect\citeauthoryear{Alvarado-G{\'o}mez et~al.,}{Alvarado-G{\'o}mez
  et~al.}{2015}]{AlvaradoGomez2015}
Alvarado-G{\'o}mez J.~D.,  et~al., 2015, Astronomy and Astrophysics, 582, A38

\bibitem[\protect\citeauthoryear{{Ammons}, {Robinson}, {Strader}, {Laughlin},
  {Fischer}  \& {Wolf}}{{Ammons} et~al.}{2006}]{Ammons2008}
{Ammons} S.~M.,  {Robinson} S.~E.,  {Strader} J.,  {Laughlin} G.,  {Fischer}
  D.,   {Wolf} A.,  2006, \mn@doi [\apj] {10.1086/498490}, \href
  {https://ui.adsabs.harvard.edu/#abs/2006ApJ...638.1004A} {638, 1004}

\bibitem[\protect\citeauthoryear{{Baraffe}, {Homeier}, {Allard}  \&
  {Chabrier}}{{Baraffe} et~al.}{2015}]{Baraffe2015}
{Baraffe} I.,  {Homeier} D.,  {Allard} F.,   {Chabrier} G.,  2015, \mn@doi
  [\aap] {10.1051/0004-6361/201425481}, \href
  {http://adsabs.harvard.edu/abs/2015A%26A...577A..42B} {577, A42}

\bibitem[\protect\citeauthoryear{{Butler}, {Marcy}, {Williams}, {McCarthy},
  {Dosanjh}  \& {Vogt}}{{Butler} et~al.}{1996}]{Butler1996}
{Butler} R.~P.,  {Marcy} G.~W.,  {Williams} E.,  {McCarthy} C.,  {Dosanjh} P.,
   {Vogt} S.~S.,  1996, \mn@doi [\pasp] {10.1086/133755}, \href
  {http://adsabs.harvard.edu/abs/1996PASP..108..500B} {108, 500}

\bibitem[\protect\citeauthoryear{Choquet et~al.,}{Choquet
  et~al.}{2016}]{Choquet2016}
Choquet {\'{E}}.,  et~al., 2016, \mn@doi [The Astrophysical Journal]
  {10.3847/2041-8205/817/1/L2}, 817, L2

\bibitem[\protect\citeauthoryear{{Collier Cameron}}{{Collier
  Cameron}}{1997}]{CollierCameron1997}
{Collier Cameron} A.,  1997, \mn@doi [\mnras] {10.1093/mnras/287.3.556}, \href
  {http://adsabs.harvard.edu/abs/1997MNRAS.287..556C} {287, 556}

\bibitem[\protect\citeauthoryear{Donati, Semel, Carter, Rees  \&
  Collier~Cameron}{Donati et~al.}{1997}]{Donati1997}
Donati J.~F.,  Semel M.,  Carter B.~D.,  Rees D.~E.,   Collier~Cameron A.,
  1997, \mnras, 291, 658

\bibitem[\protect\citeauthoryear{{Donati}, {Wade}, {Babel}, {Henrichs}, {de
  Jong}  \& {Harries}}{{Donati} et~al.}{2001}]{Donati2001}
{Donati} J.~F.,  {Wade} G.~A.,  {Babel} J.,  {Henrichs} H.~f.,  {de Jong}
  J.~A.,   {Harries} T.~J.,  2001, \mn@doi [\mnras]
  {10.1111/j.1365-2966.2001.04713.x}, \href
  {https://ui.adsabs.harvard.edu/abs/2001MNRAS.326.1265D} {326, 1265}

\bibitem[\protect\citeauthoryear{{Donati} et~al.,}{{Donati}
  et~al.}{2007}]{Donati2007}
{Donati} J.-F.,  et~al., 2007, \mn@doi [\mnras]
  {10.1111/j.1365-2966.2007.12194.x}, \href
  {http://adsabs.harvard.edu/abs/2007MNRAS.380.1297D} {380, 1297}

\bibitem[\protect\citeauthoryear{{Donati} et~al.,}{{Donati}
  et~al.}{2008}]{Donati2008}
{Donati} J.-F.,  et~al., 2008, \mn@doi [\mnras]
  {10.1111/j.1365-2966.2008.13111.x}, \href
  {http://adsabs.harvard.edu/abs/2008MNRAS.386.1234D} {386, 1234}

\bibitem[\protect\citeauthoryear{Donati et~al.,}{Donati
  et~al.}{2010a}]{Donati2010}
Donati J.~F.,  et~al., 2010a, \mn@doi [Monthly Notices of the Royal
  Astronomical Society] {10.1111/j.1365-2966.2009.15998.x}, 402, 1426

\bibitem[\protect\citeauthoryear{{Donati} et~al.,}{{Donati}
  et~al.}{2010b}]{Donati2010b}
{Donati} J.-F.,  et~al., 2010b, \mn@doi [\mnras]
  {10.1111/j.1365-2966.2010.17409.x}, \href
  {http://adsabs.harvard.edu/abs/2010MNRAS.409.1347D} {409, 1347}

\bibitem[\protect\citeauthoryear{{Donati} et~al.,}{{Donati}
  et~al.}{2011a}]{Donati2011}
{Donati} J.-F.,  et~al., 2011a, \mn@doi [\mnras]
  {10.1111/j.1365-2966.2011.19366.x}, \href
  {http://adsabs.harvard.edu/abs/2011MNRAS.417.1747D} {417, 1747}

\bibitem[\protect\citeauthoryear{Donati et~al.,}{Donati
  et~al.}{2011b}]{Donati2011b}
Donati J.~F.,  et~al., 2011b, \mn@doi [Monthly Notices of the Royal
  Astronomical Society] {10.1111/j.1365-2966.2011.19288.x}, 417, 472

\bibitem[\protect\citeauthoryear{Donati et~al.,}{Donati
  et~al.}{2012}]{Donati2012}
Donati J.-F.,  et~al., 2012, \mn@doi [Mon. Not. R. Astron. Soc]
  {10.1111/j.1365-2966.2012.21482.x}, 425, 2948

\bibitem[\protect\citeauthoryear{Donati et~al.,}{Donati
  et~al.}{2013}]{Donati2013}
Donati J.~F.,  et~al., 2013, \mn@doi [Monthly Notices of the Royal Astronomical
  Society] {10.1093/mnras/stt1622}, 436, 881

\bibitem[\protect\citeauthoryear{Donati et~al.,}{Donati
  et~al.}{2014}]{Donati2014}
Donati J.~F.,  et~al., 2014, Monthly Notices of the Royal Astronomical Society,
  444, 3220

\bibitem[\protect\citeauthoryear{Donati et~al.,}{Donati
  et~al.}{2015}]{Donati2015}
Donati J.~F.,  et~al., 2015, \mnras, 453, 3706

\bibitem[\protect\citeauthoryear{{Donati} et~al.,}{{Donati}
  et~al.}{2020}]{Donati2020}
{Donati} J.~F.,  et~al., 2020, \mn@doi [\mnras] {10.1093/mnras/staa2569}, \href
  {https://ui.adsabs.harvard.edu/abs/2020MNRAS.tmp.2502D} {}

\bibitem[\protect\citeauthoryear{Emeriau-Viard \& Brun}{Emeriau-Viard \&
  Brun}{2017}]{EmeriauViard2017}
Emeriau-Viard C.,  Brun A.~S.,  2017, \mn@doi [The Astrophysical Journal]
  {10.3847/1538-4357/aa7b33}, 846, 8

\bibitem[\protect\citeauthoryear{{Follert} et~al.,}{{Follert}
  et~al.}{2014}]{Follert2014}
{Follert} R.,  et~al., 2014, in Ground-based and Airborne Instrumentation for
  Astronomy V. p. 914719, \mn@doi{10.1117/12.2054197}

\bibitem[\protect\citeauthoryear{{Folsom} et~al.,}{{Folsom}
  et~al.}{2018}]{Folsom2018}
{Folsom} C.~P.,  et~al., 2018, \mn@doi [\mnras] {10.1093/mnras/stx3021}, \href
  {http://adsabs.harvard.edu/abs/2018MNRAS.474.4956F} {474, 4956}

\bibitem[\protect\citeauthoryear{{Gaia Collaboration} et~al.,}{{Gaia
  Collaboration} et~al.}{2016}]{GaiaCollab2016}
{Gaia Collaboration} et~al., 2016, \mn@doi [\aap]
  {10.1051/0004-6361/201629272}, \href
  {https://ui.adsabs.harvard.edu/#abs/2016A&A...595A...1G} {595, A1}

\bibitem[\protect\citeauthoryear{{Gaia Collaboration}, {Brown}, {Vallenari},
  {Prusti}, {de Bruijne}, {Babusiaux}  \& {Bailer-Jones}}{{Gaia Collaboration}
  et~al.}{2018}]{GaiaDR22018}
{Gaia Collaboration} {Brown} A.~G.~A.,  {Vallenari} A.,  {Prusti} T.,  {de
  Bruijne} J.~H.~J.,  {Babusiaux} C.,   {Bailer-Jones} C.~A.~L.,  2018,
  preprint, \href {https://ui.adsabs.harvard.edu/#abs/2018arXiv180409365G} {p.
  arXiv:1804.09365} (\mn@eprint {arXiv} {1804.09365})

\bibitem[\protect\citeauthoryear{{Gomes da Silva}, Santos, Boisse, Dumusque  \&
  Lovis}{{Gomes da Silva} et~al.}{2014}]{GomesdaSilva2014}
{Gomes da Silva} J.,  Santos N.~C.,  Boisse I.,  Dumusque X.,   Lovis C.,
  2014, \mn@doi [Astronomy {\&} Astrophysics] {10.1051/0004-6361/201322697},
  566, A66

\bibitem[\protect\citeauthoryear{Gregory, Donati, Morin, Hussain, Mayne,
  Hillenbrand  \& Jardine}{Gregory et~al.}{2012}]{Gregory2012}
Gregory S.~G.,  Donati J.-F.,  Morin J.,  Hussain G. A.~J.,  Mayne N.~J.,
  Hillenbrand L.~A.,   Jardine M.,  2012, \mn@doi [apj]
  {10.1088/0004-637X/755/2/97}, 755, 97

\bibitem[\protect\citeauthoryear{Grunhut et~al.,}{Grunhut
  et~al.}{2013}]{Grunhut2013}
Grunhut J.~H.,  et~al., 2013, \mn@doi [Monthly Notices of the Royal
  Astronomical Society] {10.1093/mnras/sts153}, 428, 1686

\bibitem[\protect\citeauthoryear{{Gully-Santiago} et~al.,}{{Gully-Santiago}
  et~al.}{2017}]{Gully2017}
{Gully-Santiago} M.~A.,  et~al., 2017, \mn@doi [\apj]
  {10.3847/1538-4357/836/2/200}, \href
  {https://ui.adsabs.harvard.edu/abs/2017ApJ...836..200G} {836, 200}

\bibitem[\protect\citeauthoryear{H{\'e}brard, Donati, Delfosse, Morin, Moutou
  \& Boisse}{H{\'e}brard et~al.}{2016}]{Hebrard2016}
H{\'e}brard {\'E}.~M.,  Donati J.~F.,  Delfosse X.,  Morin J.,  Moutou C.,
  Boisse I.,  2016, \mnras, 461, 1465

\bibitem[\protect\citeauthoryear{{Henden}, {Templeton}, {Terrell}, {Smith},
  {Levine}  \& {Welch}}{{Henden} et~al.}{2016}]{Henden2016}
{Henden} A.~A.,  {Templeton} M.,  {Terrell} D.,  {Smith} T.~C.,  {Levine} S.,
  {Welch} D.,  2016, VizieR Online Data Catalog, \href
  {http://cdsads.u-strasbg.fr/abs/2016yCat.2336....0H} {2336}

\bibitem[\protect\citeauthoryear{{Hill}, {Carmona}, {Donati}, {Hussain},
  {Gregory}, {Alencar}, {Bouvier}  \& {The Matysse Collaboration}}{{Hill}
  et~al.}{2017}]{Hill2017}
{Hill} C.~A.,  {Carmona} A.,  {Donati} J.~F.,  {Hussain} G.~A.~J.,  {Gregory}
  S.~G.,  {Alencar} S.~H.~P.,  {Bouvier} J.,   {The Matysse Collaboration}
  2017, \mn@doi [\mnras] {10.1093/mnras/stx2042}, \href
  {https://ui.adsabs.harvard.edu/#abs/2017MNRAS.472.1716H} {472, 1716}

\bibitem[\protect\citeauthoryear{{Hill}, {Folsom}, {Donati}, {Herczeg},
  {Hussain}, {Alencar}, {Gregory}  \& {Matysse Collaboration}}{{Hill}
  et~al.}{2019}]{Hill2019}
{Hill} C.~A.,  {Folsom} C.~P.,  {Donati} J.~F.,  {Herczeg} G.~J.,  {Hussain}
  G.~A.~J.,  {Alencar} S.~H.~P.,  {Gregory} S.~G.,   {Matysse Collaboration}
  2019, \mn@doi [\mnras] {10.1093/mnras/stz403}, \href
  {https://ui.adsabs.harvard.edu/abs/2019MNRAS.484.5810H} {484, 5810}

\bibitem[\protect\citeauthoryear{{Hussain}, {Donati}, {Collier Cameron}  \&
  {Barnes}}{{Hussain} et~al.}{2000}]{Hussain2000}
{Hussain} G.~A.~J.,  {Donati} J.-F.,  {Collier Cameron} A.,   {Barnes} J.~R.,
  2000, \mn@doi [\mnras] {10.1046/j.1365-8711.2000.03573.x}, \href
  {http://ukads.nottingham.ac.uk/abs/2000MNRAS.318..961H} {318, 961}

\bibitem[\protect\citeauthoryear{{Hussain}, {van Ballegooijen}, {Jardine}  \&
  {Collier Cameron}}{{Hussain} et~al.}{2002}]{Hussain2002}
{Hussain} G.~A.~J.,  {van Ballegooijen} A.~A.,  {Jardine} M.,   {Collier
  Cameron} A.,  2002, \mn@doi [\apj] {10.1086/341429}, \href
  {https://ui.adsabs.harvard.edu/abs/2002ApJ...575.1078H} {575, 1078}

\bibitem[\protect\citeauthoryear{{Hussain} et~al.,}{{Hussain}
  et~al.}{2009}]{Hussain2009}
{Hussain} G.~A.~J.,  et~al., 2009, \mn@doi [\mnras]
  {10.1111/j.1365-2966.2009.14881.x}, \href
  {http://adsabs.harvard.edu/abs/2009MNRAS.398..189H} {398, 189}

\bibitem[\protect\citeauthoryear{{Hussain} et~al.,}{{Hussain}
  et~al.}{2016}]{Hussain2016}
{Hussain} G.~A.~J.,  et~al., 2016, \mn@doi [\aap]
  {10.1051/0004-6361/201526595}, \href
  {http://adsabs.harvard.edu/abs/2016A%26A...585A..77H} {585, A77}

\bibitem[\protect\citeauthoryear{{Low}, {Smith}, {Werner}, {Chen}, {Krause},
  {Jura}  \& {Hines}}{{Low} et~al.}{2005}]{Low2005}
{Low} F.~J.,  {Smith} P.~S.,  {Werner} M.,  {Chen} C.,  {Krause} V.,  {Jura}
  M.,   {Hines} D.~C.,  2005, \mn@doi [\apj] {10.1086/432640}, \href
  {https://ui.adsabs.harvard.edu/abs/2005ApJ...631.1170L} {631, 1170}

\bibitem[\protect\citeauthoryear{Marsden et~al.,}{Marsden
  et~al.}{2014}]{Marsden2014}
Marsden S.~C.,  et~al., 2014, \mn@doi [\mnras] {10.1093/mnras/stu1663}, 444,
  3517

\bibitem[\protect\citeauthoryear{{McDonald}, {Zijlstra}  \&
  {Watson}}{{McDonald} et~al.}{2017}]{McDonald2017}
{McDonald} I.,  {Zijlstra} A.~A.,   {Watson} R.~A.,  2017, \mn@doi [\mnras]
  {10.1093/mnras/stx1433}, \href
  {https://ui.adsabs.harvard.edu/#abs/2017MNRAS.471..770M} {471, 770}

\bibitem[\protect\citeauthoryear{Mentuch, Brandeker, van Kerkwijk, Jayawardhana
   \& Hauschildt}{Mentuch et~al.}{2008}]{Mentuch2008}
Mentuch E.,  Brandeker A.,  van Kerkwijk M.,  Jayawardhana R.,   Hauschildt P.,
   2008, \mn@doi [\apj] {10.1086/592764}, 689, 1127

\bibitem[\protect\citeauthoryear{Nicholson, Hussain, Donati, Folsom, Mengel,
  Carter  \& Wright}{Nicholson et~al.}{2018}]{Nicholson2018}
Nicholson B.~A.,  Hussain G. A.~J.,  Donati J.-F.,  Folsom C.~P.,  Mengel M.,
  Carter B.~D.,   Wright D.,  2018, \mn@doi [Monthly Notices of the Royal
  Astronomical Society] {10.1093/mnras/sty1965}, 480, 1754

\bibitem[\protect\citeauthoryear{{Olofsson} et~al.,}{{Olofsson}
  et~al.}{2018}]{Olofsson2018}
{Olofsson} J.,  et~al., 2018, preprint, \href
  {http://adsabs.harvard.edu/abs/2018arXiv180401929O} {} (\mn@eprint {arXiv}
  {1804.01929})

\bibitem[\protect\citeauthoryear{Pecaut \& Mamajek}{Pecaut \&
  Mamajek}{2013}]{Pecaut2013}
Pecaut M.~J.,  Mamajek E.~E.,  2013, The Astrophysical Journal Supplement
  Series, 208, 9

\bibitem[\protect\citeauthoryear{{Ryabchikova}, {Piskunov}, {Kurucz},
  {Stempels}, {Heiter}, {Pakhomov}  \& {Barklem}}{{Ryabchikova}
  et~al.}{2015}]{Ryabchikova2015}
{Ryabchikova} T.,  {Piskunov} N.,  {Kurucz} R.~L.,  {Stempels} H.~C.,  {Heiter}
  U.,  {Pakhomov} Y.,   {Barklem} P.~S.,  2015, \mn@doi [\physscr]
  {10.1088/0031-8949/90/5/054005}, \href
  {http://adsabs.harvard.edu/abs/2015PhyS...90e4005R} {90, 054005}

\bibitem[\protect\citeauthoryear{{Setiawan}, {Henning}, {Launhardt},
  {M{\"u}ller}, {Weise}  \& {K{\"u}rster}}{{Setiawan}
  et~al.}{2008}]{setiawan2008}
{Setiawan} J.,  {Henning} T.,  {Launhardt} R.,  {M{\"u}ller} A.,  {Weise} P.,
  {K{\"u}rster} M.,  2008, \mn@doi [\nat] {10.1038/nature06426}, \href
  {https://ui.adsabs.harvard.edu/abs/2008Natur.451...38S} {451, 38}

\bibitem[\protect\citeauthoryear{Siess, Dufour  \& Forestini}{Siess
  et~al.}{2000}]{Siess2000}
Siess L.,  Dufour E.,   Forestini M.,  2000, Astronomy and Astrophysics, 358,
  593

\bibitem[\protect\citeauthoryear{Simitev \& Busse}{Simitev \&
  Busse}{2009}]{Simitev2009}
Simitev R.~D.,  Busse F.~H.,  2009, EPL (Europhysics Letters), 85, 19001

\bibitem[\protect\citeauthoryear{{Song}, {Zuckerman}  \& {Bessell}}{{Song}
  et~al.}{2003}]{Song2003}
{Song} I.,  {Zuckerman} B.,   {Bessell} M.~S.,  2003, \mn@doi [The
  Astrophysical Journal] {10.1086/379194}, \href
  {http://adsabs.harvard.edu/abs/2003ApJ...599..342S} {599, 342}

\bibitem[\protect\citeauthoryear{{Valenti} \& {Fischer}}{{Valenti} \&
  {Fischer}}{2005}]{Valenti2005}
{Valenti} J.~A.,  {Fischer} D.~A.,  2005, \mn@doi [\apjs] {10.1086/430500},
  \href {http://adsabs.harvard.edu/abs/2005ApJS..159..141V} {159, 141}

\bibitem[\protect\citeauthoryear{{Wade}, {Donati}, {Landstreet}  \&
  {Shorlin}}{{Wade} et~al.}{2000}]{Wade2000}
{Wade} G.~A.,  {Donati} J.~F.,  {Landstreet} J.~D.,   {Shorlin} S.~L.~S.,
  2000, \mn@doi [\mnras] {10.1046/j.1365-8711.2000.03271.x}, \href
  {https://ui.adsabs.harvard.edu/abs/2000MNRAS.313..851W} {313, 851}

\bibitem[\protect\citeauthoryear{Webb, Zuckerman, Platais, Patience, White,
  Schwartz  \& McCarthy}{Webb et~al.}{1999}]{Webb1999}
Webb R.~A.,  Zuckerman B.,  Platais I.,  Patience J.,  White R.~J.,  Schwartz
  M.~J.,   McCarthy C.,  1999, \mn@doi [The Astrophysical Journal]
  {10.1086/311856}, 512, L63

\bibitem[\protect\citeauthoryear{{Wittenmyer} \& {Marshall}}{{Wittenmyer} \&
  {Marshall}}{2015}]{debris}
{Wittenmyer} R.~A.,  {Marshall} J.~P.,  2015, \mn@doi [\aj]
  {10.1088/0004-6256/149/2/86}, \href
  {http://adsabs.harvard.edu/abs/2015AJ....149...86W} {149, 86}

\bibitem[\protect\citeauthoryear{{Wittenmyer}, {Endl}, {Cochran}, {Hatzes},
  {Walker}, {Yang}  \& {Paulson}}{{Wittenmyer} et~al.}{2006}]{limitspaper}
{Wittenmyer} R.~A.,  {Endl} M.,  {Cochran} W.~D.,  {Hatzes} A.~P.,  {Walker}
  G.~A.~H.,  {Yang} S.~L.~S.,   {Paulson} D.~B.,  2006, \mn@doi [\aj]
  {10.1086/504942}, \href {http://adsabs.harvard.edu/abs/2006AJ....132..177W}
  {132, 177}

\bibitem[\protect\citeauthoryear{{Wittenmyer} et~al.,}{{Wittenmyer}
  et~al.}{2016}]{newjupiters}
{Wittenmyer} R.~A.,  et~al., 2016, \mn@doi [\apj] {10.3847/0004-637X/819/1/28},
  \href {http://adsabs.harvard.edu/abs/2016ApJ...819...28W} {819, 28}

\bibitem[\protect\citeauthoryear{{Wittenmyer} et~al.,}{{Wittenmyer}
  et~al.}{2020}]{ppps8}
{Wittenmyer} R.~A.,  et~al., 2020, \mn@doi [\mnras] {10.1093/mnras/stz3378},
  \href {https://ui.adsabs.harvard.edu/abs/2020MNRAS.491.5248W} {491, 5248}

\bibitem[\protect\citeauthoryear{Yang, Johns-Krull  \& Valenti}{Yang
  et~al.}{2008}]{Yang2008}
Yang H.,  Johns-Krull C.~M.,   Valenti J.~A.,  2008, \mn@doi [The Astronomical
  Journal] {10.1088/0004-6256/136/6/2286}, 136, 2286

\bibitem[\protect\citeauthoryear{Yu et~al.,}{Yu et~al.}{2017}]{Yu2017}
Yu L.,  et~al., 2017, \mn@doi [\mnras] {10.1093/mnras/stx009}, 467, 1342

\bibitem[\protect\citeauthoryear{{Yu} et~al.,}{{Yu} et~al.}{2019}]{Yu2019}
{Yu} L.,  et~al., 2019, \mn@doi [\mnras] {10.1093/mnras/stz2481}, \href
  {https://ui.adsabs.harvard.edu/abs/2019MNRAS.489.5556Y} {489, 5556}

\bibitem[\protect\citeauthoryear{{Zechmeister} \& {K{\"u}rster}}{{Zechmeister}
  \& {K{\"u}rster}}{2009}]{Zechmeister2009}
{Zechmeister} M.,  {K{\"u}rster} M.,  2009, \mn@doi [\aap]
  {10.1051/0004-6361:200811296}, \href
  {http://adsabs.harvard.edu/abs/2009A%26A...496..577Z} {496, 577}

\bibitem[\protect\citeauthoryear{{da Silva}, {Torres}, {de La Reza}, {Quast},
  {Melo}  \& {Sterzik}}{{da Silva} et~al.}{2009}]{daSilva2009}
{da Silva} L.,  {Torres} C.~A.~O.,  {de La Reza} R.,  {Quast} G.~R.,  {Melo}
  C.~H.~F.,   {Sterzik} M.~F.,  2009, \mn@doi [\aap]
  {10.1051/0004-6361/200911736}, \href
  {http://adsabs.harvard.edu/abs/2009A%26A...508..833D} {508, 833}

\makeatother
\end{thebibliography}

\bsp	
\label{lastpage}
\end{document}